\documentclass[journal]{vgtc}                
\usepackage[dvipsnames]{xcolor}
\usepackage{amsmath,amsfonts,amssymb}
\usepackage[english]{babel}    
\usepackage{chngcntr}          

\usepackage{comment}
\usepackage{gensymb}  
\usepackage{booktabs} 
\usepackage{siunitx} 

\usepackage{enumitem}

\ieeedoi{10.1109/TVCG.2019.2934536}

\addto{\captionsenglish}{}

\newcommand{\figskip}{\vspace{-3mm}}

\DeclareMathAlphabet{\mathsfit}{T1}{\sfdefault}{\mddefault}{\sldefault} 

\ifpdf
  \pdfoutput=1\relax                   
  \usepackage{graphicx}                
  \DeclareGraphicsExtensions{.pdf,.png,.jpg,.jpeg} 
\else
  \usepackage{graphicx}                
  \DeclareGraphicsExtensions{.eps}     
\fi%

\graphicspath{{figures/}{pictures/}{images/}{./}} 

\usepackage{microtype}                 
\PassOptionsToPackage{warn}{textcomp}  
\usepackage{textcomp}                  
\usepackage{mathptmx}                  
\usepackage{times}                     
\usepackage{cite}                      
\usepackage{tabu}                      
\usepackage{booktabs}                  

\onlineid{1119}

\vgtccategory{Research}
\vgtcpapertype{Evaluation}

\title{Semantic Discriminability for Visual Communication}

\author{Karen B. Schloss, Zachary Leggon, and Laurent Lessard}

\authorfooter{
\item  Karen B. Schloss, Psychology and Wisconsin Institute for Discovery, University of Wisconsin--Madison. Email: kschloss@wisc.edu.
\item Zachary Leggon, Biology and Wisconsin Institute for Discovery, University of Wisconsin--Madison,  Email: zleggon@wisc.edu.
\item  Laurent Lessard, Mechanical and Industrial Engineering, Northeastern University. Email: l.lessard@northeastern.edu.
}

\shortauthortitle{Schloss \MakeLowercase{\textit{et~al.}}: Semantic Discriminability for Visual Communication}

\abstract{To interpret information visualizations, observers must determine how visual features map onto concepts. First and foremost, this ability depends on perceptual discriminability; e.g., observers must be able to see the difference between different colors for those colors to communicate different meanings. However, the ability to interpret visualizations also depends on semantic discriminability, the degree to which observers can infer a unique mapping between visual features and concepts, based on the visual features and concepts alone (i.e., without help from verbal cues such as legends or labels).  Previous evidence suggested that observers were better at interpreting encoding systems that maximized semantic discriminability (maximizing association strength between assigned colors and concepts while minimizing association strength between unassigned colors and concepts), compared to a system that only maximized color-concept association strength. However, increasing semantic discriminability also resulted in increased perceptual distance, so it is unclear which factor was responsible for improved performance. In the present study, we conducted two experiments that tested for independent effects of semantic distance and perceptual distance on semantic discriminability of bar graph data visualizations. Perceptual distance was large enough to ensure colors were more than just noticeably different. We found that increasing semantic distance improved performance, independent of variation in perceptual distance, and when these two factors were uncorrelated, responses were dominated by semantic distance. These results have implications for navigating trade-offs in color palette design optimization for visual communication.}


\CCScatlist{ 
 \CCScat{K.6.1}{Management of Computing and Information Systems}%
{Project and People Management}{Life Cycle};
 \CCScat{K.7.m}{The Computing Profession}{Miscellaneous}{Ethics}
}

\teaser{
  \centering
  \includegraphics[width=\linewidth]{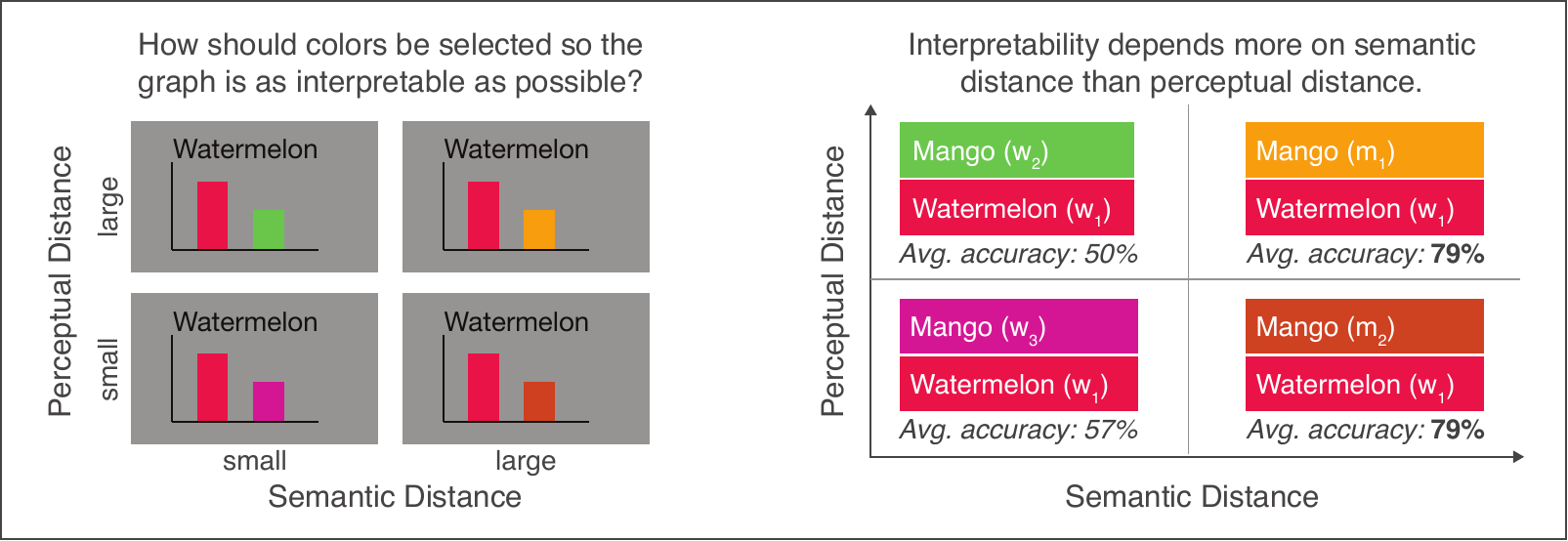}
  \caption{This study investigated how varying perceptual and semantic distance influenced people's ability to interpret color palettes.
  Left: four example trials from Experiment~2, in which participants reported which bar (left/right) corresponded to the target fruit named above the graph (either mango or watermelon).
  No legend was provided, so participants had to rely on their inferred mappings to complete the task.
  Right: aggregated results corresponding to the color pairs from the left panel. The colors are labeled with their correct assignments (see Section~\ref{sec:defInterp} for details) as well as the color names in parentheses (see Table \ref{table:ExpColors} for coordinates). Participants had higher mean accuracy when the \textit{semantic distance} between the pair of colors was greater (a notion we define in Section~\ref{sec:semantic_distance}), whereas perceptual discriminability had little effect on accuracy when controlling for semantic distance.
  }\label{fig:Overview}
}

 \nocopyrightspace

\vgtcinsertpkg

\begin{document}


\firstsection{Introduction}
\maketitle

In visual communication, designers produce information visualizations by encoding concepts in visual features, and observers interpret visualizations by decoding concepts from visual features \cite{cleveland1987,wood1968}. Interpreting visualizations involves multiple component processes, including (1) perceiving and identifying important features within a visualization, (2) mapping those features to the concepts they represent, and (3) deriving implications about information represented in the visualization \cite{shah2002}. For example, to interpret visualizations that encode categories using color (e.g., bar graphs, choropleth maps, transit maps, and recycling bin signage), observers must perceive and distinguish the colors, determine how each color in the palette maps to a category, and then use that mapping to glean knowledge from the visualization (e.g., patterns of data from bar graphs or choropleth maps, which train to take from a transit map, or where to discard paper from recycling signage). 

Many factors influence interpretability, including (1) characteristics of visualizations, (2) observers' knowledge about visualizations, and (3) observers' knowledge about content in visualizations (outlined in \cite{shah2002}, with respect to graphs). Here, we focus on how interpretability is influenced by perceptual characteristics of visualizations that are used to encode meaning (e.g., colors) and observers' semantic associations with those perceptual features. Thus, for present purposes, we operationalize ``interpretability'' as the ability to accurately decode the encoded mapping. We aim to develop a deeper understanding of how interpretability is influenced by two properties:  \textit{perceptual discriminability} and \textit{semantic discriminability}. 

\subsection{Perceptual discriminability} Perceptual discriminability is the degree to which observers can perceive differences between different visual features (e.g., colors, sizes, shapes, or textures). Some amount of perceptual discriminability is necessary because observers cannot decode different meanings from perceptually identical features \cite{kosslyn2006} (e.g., they must be able to perceive the difference between two shades of blue to decode that those blues encode different concepts). Thus, visualization research has emphasized the importance of understanding perceptual discriminability \cite{healey1996, stone2014, szafir2018, gramazio2017}, including how it varies with mark size \cite{stone2014} and shape \cite{szafir2018}. And, design guidelines have emphasized the importance of representing categorical information with colors that are well-separated in color space \cite{shah2002, healey1996, harrower2003}. If perceptually discriminable features are accompanied by verbal descriptions specifying the encoded mapping (e.g., legends or labels), then observers have all the information required to decode the encoded mapping. This rationale supports using pre-made color palettes (e.g., Tableau and Colorbrewer palettes \cite{harrower2003}) that have been designed to ensure perceptual discriminability. However, the ability to decode encoded mappings depends on more than perceptual discriminability and legend reading, as explained below. 

\subsection{Semantic discriminability} \label{sec:SemDisc}

We define semantic discriminability as the degree to which observers can infer a unique mapping between visual features and concepts, based on the visual features and concepts alone (i.e., without legends or labels). For example, if observers are given an unlabeled graph containing yellow and blue colored bars and are told the graph is about the concepts banana and blueberry, they could easily infer that yellow maps to banana and blue maps to blueberry. This is because yellow and blue are semantically discriminable for the concepts banana and blueberry. Conversely, it would be more difficult to infer how orangish-yellow and greenish-yellow map to the concepts banana and lemon because both colors are similarly associated with both concepts, and thus less semantically discriminable. 

Semantic discriminability might sound similar to interpretability, but they are distinct constructs. Semantic discriminability concerns the ability to infer a unique mapping (irrespective of the encoded mapping), whereas interpretability concerns the ability to decode the correct mapping (specified by the encoded mapping). Building on the banana/blueberry graph example, yellow and blue would be semantically discriminable colors, regardless of the encoded mapping in the graph. Observers would infer that yellow maps to banana and blueberry maps to blue. Now, if the encoded mapping was yellow-banana/blue-blueberry, the graph would be easy to interpret because the encoded mapping matched the inferred mapping. But, if the encoded mapping was blue-banana/yellow-blueberry (i.e., cross-mapped \cite{gentner1986}), the graph would be harder to interpret because the encoded mapping did not match the inferred mapping (i.e., Kosslyn's compatibility principle \cite{kosslyn2006}, Tversky et al.'s congruence principle \cite{tversky2002}). Observers are better at interpreting colors in visualizations when encoded mappings match inferred mappings, even when there is a clear legend \cite{lin2013, schloss2019}.

Based on the examples above, one might conclude that interpretability depends only on association strengths of encoded color-concept pairs. Lin et al.~\cite{lin2013} referred to palettes in which colors evoke the concepts they represent as \textit{semantically resonant} color palettes. However, interpretability can also be achieved when not all color-concept pairs are semantically resonant \cite{schloss2018}, see Section \ref{sec:asigninf}. Rathore et al.~\cite{rathore2020} referred to this more general case as \textit{semantically interpretable} color palettes. For simplicity, we use the term \textit{interpretability} in the present work to refer to the more general case.

\subsection{Perceptual vs.\ semantic discriminability?} 
From prior work, it is clear that interpretability hinges on some degree of perceptual discriminability \cite{stone2014, szafir2018, healey1996, brewer1994} and interpretability benefits from semantic discriminability \cite{schloss2018}. However, given previous research, it is currently unknown whether increasing semantic discriminability improves interpretability, beyond that which can be explained by perceptual discriminability. Returning to our banana/blueberry/lemon examples used so far in this introduction, these examples were intended to build the intuition for semantic discriminability, but they confounded perceptual and semantic discriminability. Yellow and blue are both high in perceptual and semantic discriminability when encoding for the concepts banana and blueberry, and orangish-yellow vs.\ greenish-yellow are both low in perceptual and semantic discriminability when encoding for the concepts banana and lemon (assuming trichromatic color vision).
Similarly, in prior work that suggested semantic discriminability improved interpretability, colors in the more semantically discriminable color palette were closer together in color space \cite{schloss2018}, see Section \ref{sec:asigninf}. So, it is unclear if this improvement was due to semantic or perceptual discriminability. Yet, semantic and perceptual discriminability \emph{can} vary independently (Fig.~\ref{fig:Overview}), and understanding their independent effects on interpretability is important for determining how to resolve conflicts between them when optimizing color palette design.

Likewise, it is also unknown whether increasing perceptual discriminability beyond that which is needed for semantic discriminability influences interpretability.  For two colors to be semantically discriminable, they must be sufficiently perceptually discriminable to tell them apart. Otherwise, observers could not reliably infer that one color maps more than another color does to a given concept. Thus, perceptual discriminability might not capture additional variance in interpretability beyond that which is explained by semantic discriminability. 
 
To address these questions, we tested for independent effects of perceptual and semantic discriminability on interpretability. The results will not only provide a deeper understanding about the relative contributions of perceptual and cognitive factors for visual reasoning, but will also inform optimal color palette design. Designing effective palettes for information visualization requires navigating trade-offs between several, sometimes competing, factors (e.g., perceptual discriminability, semantic discriminability, name difference, emotional connotation, and aesthetics) \cite{heer2009, stone2014, gramazio2017, bartram2017, lin2013}. Understanding the relative contribution of semantic and perceptual discriminability for interpretability will inform how to prioritize these factors when conflicts arise. 

\textbf{Contributions.} Our study makes the following contributions. First, we define a metric called \textit{semantic distance} for operationalizing semantic discriminability. Semantic distance depends on the relative association strengths between each color and each concept in the context of an encoding system (see Section \ref{sec:metrics}). This is unlike perceptual distance, which only depends on the appearance of the two colors. The semantic distance between a given pair of colors may be large in the context of some concepts, but small in the context of other concepts.

Second, we present the results of two experiments that assess how perceptual distance and semantic distance influence interpretability. Evidence indicates that both factors can contribute to interpretability, but semantic distance dominates when the factors conflict. The results imply that increasing perceptual distance beyond that which is needed for semantic discriminability can improve interpretability, but when in conflict, priority should be given to maximizing semantic distance. 

\section{Background} \label{sec:background}
When people interpret information visualizations, they do not simply absorb the displayed information in a bottom-up fashion. Instead, they have biases, or expectations, about how visual features map to meanings, which guide their interpretations. These biases span topics across the field of information visualization, including graphical perception \cite{zacks1999, xiong2019}, visualizing uncertainty \cite{ruginski2016}, and color \cite{cuff1973, mcgranaghan1989, lin2013, schloss2018, schloss2019}. Understanding and designing visualizations that align with these biases will help make visualizations that are easier to interpret \cite{norman2013, tversky2011, hegarty2011}. In cases where this alignment may not be possible (i.e., multiple conflicting biases relevant to a particular visualization), an understanding of when expectations are violated can guide compensatory design decisions (e.g., extra labeling or verbal descriptions of the visualization). 

Here, we focus on understanding expectations about assignments between colors and concepts for interpreting visualizations about categorical information. However, this discussion should apply to assignments between other perceptual features and concepts, as long as people have systematic associations between those features and concepts. In this section, we first describe how designers use \textit{assignment problems} to produce encoded mappings between visual features and concepts. We then present evidence that observers use assignment inference to decode encoded mappings when interpreting visualizations. 

\subsection{Assignment problems for encoding} \label{sec:assign}
\begin{figure}[ht!!]
	\centering
	\includegraphics[width=1.0\columnwidth]{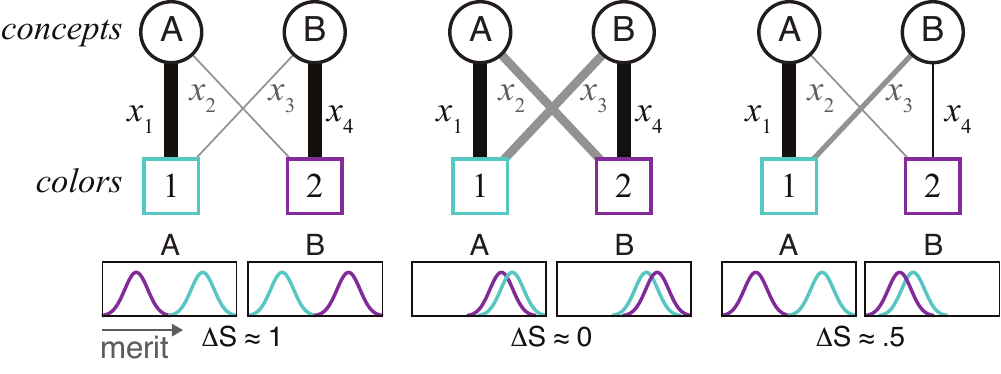}
	\vspace{-6mm}
	\caption{Bipartite graphs representing possible assignment problems between concepts (circles) and colors (squares). The \textit{merit} of each pairing is represented by the thickness of edges connecting each color to each concept. Here, the merit is color-concept association strength. Plots below each bipartite graph show the relative merit distribution of each color for each concept, assuming that association strength is a normally distributed random variable. In all three examples, mean association strengths satisfy $x_1+x_4 > x_2+x_3$, so the outcomes of the assignment problems are the same: concept A is assigned with color 1 and concept B is assigned with color 2. Black lines indicate the chosen assignment and gray lines indicate the non-chosen assignment. $\Delta S$ indicates approximate semantic distance for the examples.}
	\label{fig:Bipartite}
\end{figure}

Assignment problems have been used to create color palettes that optimize encodings between visual features and concepts \cite{lin2013, schloss2018}. An assignment problem is a model for assigning items in one category (e.g., colors) to items in another category (e.g., concepts) in a manner that maximizes a total merit score \cite{kuhn1955, munkres1957}. Assignment problems can be represented as bipartite graphs, as shown in Fig.~\ref{fig:Bipartite}. The square nodes are colors and the circular nodes are concepts. Edges are drawn between each color and each concept. The number on each edge represents the ``merit score'' of assigning that particular color to that concept, represented as $x_1,\dots,x_4$. Merit scores can be calculated using different methods \cite{lin2013, schloss2018} but the goal is always the same: construct a 1-to-1 assignment between each color and a concept, such that the sum of the merit scores of assigned color-concept pairs is maximized. 

Assignment problems are deterministic. When there are two concepts and two colors as in Fig.~\ref{fig:Bipartite}, there are only two possible outcomes and the outcome is completely determined by the merit scores on each of the four edges. If the sum of the outer edges is greater than the sum of the inner edges ($x_1+x_4 > x_2+x_3$), then concept A is assigned to color 1 and concept B is assigned to color 2. If the sum of the inner edges is greater ($x_2+x_3 > x_1+x_4$), then concept A is assigned to color 2 and concept B to color 1. Fig.~\ref{fig:Bipartite} illustrates bipartite graphs with three different patterns of merit on the edges, but all three scenarios produce identical outcomes: concept A is assigned to color 1 and concept B is assigned to color 2 because the merit scores satisfy $x_1+x_4 > x_2+x_3$. In the left and center bipartite graphs, the solution to the assignment problem (``global solution'') matches the  solution for each concept in isolation (``local solution'', concept A is more associated with color 1 than color 2, and concept B is more associated with color 2 than color 1). However, in the rightmost bipartite graph, the local and global solutions conflict: concept B is more associated with color 1, yet it is assigned to color 2. This distinction is relevant for discussing how humans decode encoded mappings (Section \ref{sec:asigninf}).

\subsection{Assignment inference for decoding} \label{sec:asigninf}
When people decode encoded mappings, they use a process similar to solving an assignment problem, called \textit{assignment inference} \cite{schloss2018}. In assignment inference, people estimate the merit of different assignments based on association strengths between visual features and concepts, and determine the assignment that maximizes merit. However, unlike how computers solve assignment problems, human assignment inference is probabilistic rather than deterministic \cite{schloss2018}. Overall, humans can produce reliable inferences, but their responses are noisy. This noise can be attributed to uncertainty in the color-concept associations that serve as input into the assignment problem. In~\cite{schloss2018}, this uncertainty was built into models that were effective at predicting human responses.

Fig.~\ref{fig:Bipartite} represents the noise in color-concept associations as distributions for each color-concept pairing. For each distribution, the mean corresponds to edge thickness in the bipartite graph. The variability is assumed to be normal. Assume each time a person does assignment inference, they draw a random value from these distributions to estimate merit for each edge. When the distributions are far apart (Fig.~\ref{fig:Bipartite} left), random draws will consistently result in the same outcome of the assignment problem. However, when distributions overlap (Fig.~\ref{fig:Bipartite} middle), random draws can result in different outcomes of the assignment problem, which results in more uncertainty in assignment inference \cite{schloss2018}. This is the basis for our semantic distance metric in the present work (see Section \ref{sec:semantic_distance}).

Evidence suggests that people perform global assignment inference when interpreting the meanings of colors in information visualizations \cite{schloss2018}. In some cases, the global solution conflicts with local solutions (Fig.~\ref{fig:Bipartite}, right). Such conflicts can result in people inferring that concepts are assigned to their most weakly associated colors, even when stronger candidate colors exist in the palette. Schloss et al.~\cite{schloss2018} first demonstrated this phenomenon using a recycling task: participants were presented with images of two colored bins, along with a word describing one ``target'' concept. There were two possible targets, paper and trash. When trash was the target and was presented with white and purple bins, participants were faced with a scenario like in Fig.~\ref{fig:Bipartite}, right. Trash was more strongly associated with white than with purple ($x_3>x_4$), but so was paper ($x_1>x_2$), and the association between paper and white was especially strong. Participants reliably discarded trash into the purple bin, even though trash was more strongly associated with white (analogous to blue in Fig.~\ref{fig:Bipartite}, right). 

 Schloss et al.~\cite{schloss2018} also assessed methods for calculating merit to generate the encoded mapping, one that maximized association strength (isolated merit function) and one that prioritized semantic discriminability over association strength (balanced merit function, although the term semantic discriminability was not used in \cite{schloss2018}). Within both color palettes, responses were faster and more accurate when the target concept was more strongly associated with its correct color, but participants were more accurate for the balanced palette than the isolated palette.  These results suggest interpretability increases with semantic discriminability. However, in addition to being more semantically discriminable, colors in the balanced palette were also further apart in CIELAB space (see Fig. \ref{fig:RecColors} in the Supplementary Material of the present paper). Thus, it is unclear if colors in the balanced color palette were easier to interpret because they were more semantically discriminable, more perceptually discriminable, or both.

\begin{figure*}[ht]
	\centering
	\includegraphics[width=0.9\linewidth]{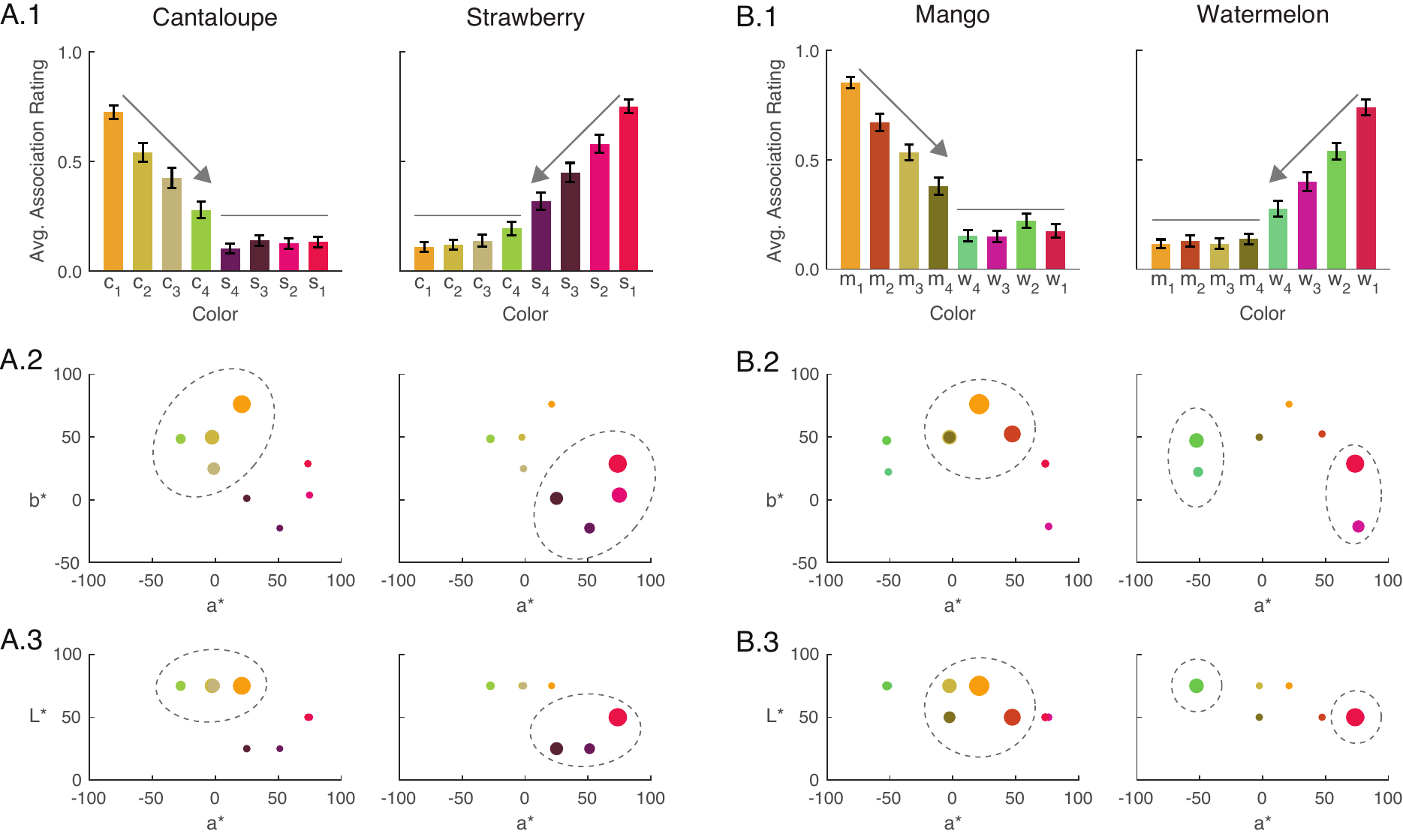}
	\vspace{-4mm}
	\caption{(A) Color-concept associations and CIELAB coordinates for colors tested in Experiment 1. (A.1) Mean association strength between each color with the concept labeled at the top of the column (error bars are standard errors of the means. Bar colors indicate the colors that were tested (see Table~\ref{table:ExpColors} for coordinates). CIELAB coordinates are shown for each color on the (A.2) a*, b* plane and the a*, L* plane (A.3), with the size of the marks corresponding to association strength with the concept named at the top of the column (same data as in A.1). Dashed ovals enclose the four colors that were the relatively strong associates with the concept at the top of the column. (B) Same as A, but for the colors and concepts in Experiment 2. Each plot has eight points, one for each color, but in some cases fewer points are visible due to occlusion on the 2D plane. }
	\label{fig:FruitColors}
	\vspace{-4mm}
\end{figure*}

\section{Approach}
In the present study, we assessed the independent effects of semantic discriminability and perceptual discriminability on participants' interpretations of bar graph data visualizations. The paradigm was the same as in Schloss et al.~\cite{schloss2018}, but instead of interpreting colors of unlabeled trash and recycling bins, participants interpreted colors of unlabeled bars in a bar graph. On each trial, participants saw a graph containing two different colored bars, along with a target fruit concept described above the graph (Fig.~\ref{fig:Overview}, left). Their task was to indicate which colored bar, left or right, corresponded to the target fruit. Within each experiment, participants judged all pairwise combinations of eight colors for two fruits (cantaloupe and strawberry in Experiment 1, mango and watermelon in Experiment 2).  The data from the present experiment and analysis code are at \url{github.com/SchlossVRL/semantic-discriminability}. 

We used a previous dataset on color-concept associations from Rathore et al.~\cite{rathore2020} to select the colors for the present study (Section \ref{sec:selecting}), define accuracy for the present tasks (Section \ref{sec:defInterp}), and quantify semantic distance (Section \ref{sec:semantic_distance}). In~\cite{rathore2020}, participants rated association strengths between each of 12 fruits and each of 58 colors (UW-58 colors), uniformly sampled in CIELAB space ($\Delta E = 25$). This distance should be at least one noticeable difference  \cite{stone2014, szafir2018}. Further details on the methods of \cite{rathore2020} are in Supplementary Material of the present paper.

\subsection{Selecting colors and concepts} \label{sec:selecting}
We chose the colors and fruit concepts using the mean color-concept association data from Rathore et al.~\cite{rathore2020} (see Table \ref{table:assocData} in the Supplementary Material) and color distances in CIELAB space ($\Delta E$). In Experiment 1, we selected eight colors and two fruits to have the following properties. Four colors varied from moderately to strongly associated with the first fruit while being weakly associated with the second fruit (Fig. \ref{fig:FruitColors}A.1, left). The other four colors varied from moderately to strongly associated with the second fruit while being weakly associated with the first fruit (Fig. \ref{fig:FruitColors}A.1, right). We generated candidate colors and fruits using an optimization routine that enforced a minimum difference in association ratings for the four colors that varied and a weak association rating for the remaining four colors. This yielded a list of candidate palettes. We excluded palettes involving fruits that had colors in their names (blueberry and orange), and this led us to selecting cantaloupe and strawberry for our first experiment. The CIELAB coordinates of these colors are plotted on the a*, b* plane in Fig.~\ref{fig:FruitColors}A.2 and on the a*, L* plane in Fig.~\ref{fig:FruitColors}A.3. The size of the marks represent the association strengths shown in \ref{fig:FruitColors}A.1. It can be seen that there are two separate clusters for ``cantaloupe colors'' and ``strawberry colors'' (indicated by the dashed ovals). In Experiment 2, we selected fruits and colors so they had the same color-concept association properties as in Experiment 1 (compare Fig.~\ref{fig:FruitColors}B.1 to \ref{fig:FruitColors}A.1). However, unlike Experiment 1, the colors in Experiment 2 are no longer clustered in CIELAB space (Fig. \ref{fig:FruitColors}B.2 and Fig. \ref{fig:FruitColors}B.3); the ``watermelon colors'' were split on either side of the ``mango colors''. These properties enabled us to independently vary semantic distance and perceptual distance.

\subsection{Quantifying metrics} \label{sec:metrics}

\subsubsection{Interpretability} \label{sec:defInterp}
We operationalized interpretability as the \textit{accuracy} of decoding the encoded mapping. The bar graphs in this study were unlabeled, so there was no explicit encoded mapping from the perspective of the participants (i.e., no objectively correct answer). However, we can determine an optimal encoded mapping by solving an assignment problem for each pair of colors and concepts and use the solution to define the ``correct'' response. The input to the assignment problem was the set of association strengths between each color-concept pair (Fig. \ref{fig:FruitColors}). Recall these data came from different participants than those in the present study. We solved the assignment problem for each pair of colors and concepts using the method described in Section \ref{sec:assign}. 

\subsubsection{Semantic discriminability}
\label{sec:semantic_distance}

We operationalized semantic discriminability using a new metric, called \textit{semantic distance} ($\Delta S$). To build the intuition for semantic distance, consider semantic discriminability in the context of assignment problems, described in Section \ref{sec:assign}. Given color-concept association ratings, the solution to the assignment problem yields a deterministic assignment of colors to concepts. However, we want to distinguish between \emph{robust} assignments (e.g., blueberry--blue and banana--yellow, which have high semantic discriminability), and \emph{fragile} assignments (e.g., 
 lemon--greenish-yellow and banana--orangish-yellow, which have low semantic discriminability since both fruits can be either color). In a robust assignment, we can expect all people to come to the same conclusion. But in a fragile assignment, people might disagree on which assignment is correct, and the same person might even respond differently when asked the same question again. We account for variability across individuals by assuming the association ratings between colors and concepts are normally distributed with a mean equal to the mean association rating and variance that is largest when the association rating is closest to the center of the rating scale.

\begin{figure}[htb]
	\centering
	\includegraphics[width=1.0\columnwidth]{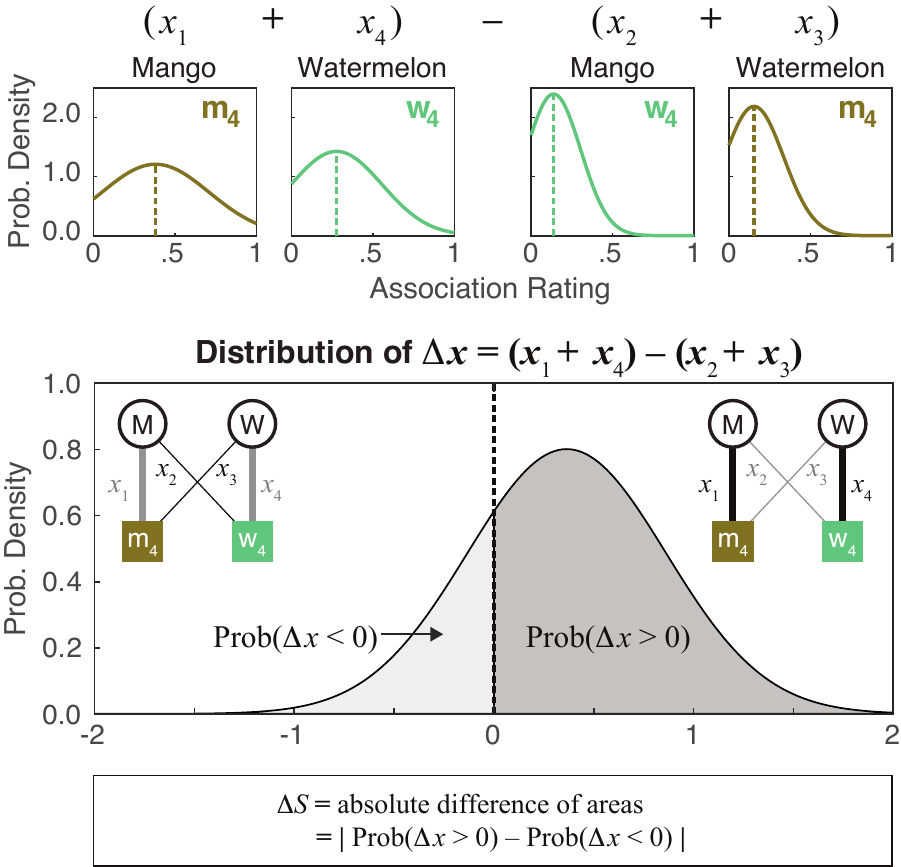} \figskip
	\caption{Illustration of semantic distance calculation using pairwise association ratings $x_i$ between the colors $m_4$ and $w_4$ and the concepts Mango (M) and Watermelon (W). Distributions are normal distributions fit to individual participants' color-concept association ratings, with the dashed line showing the mean corresponding to the bars in Fig.~\ref{fig:FruitColors}B.1. The $x_i$ are assumed to be normally distributed and combine to form $\Delta x$. When $\Delta x > 0$, the assignment M-$m_4$/W-$w_4$ is chosen and when $\Delta x < 0$, the alternative assignment M-$w_4$/W-$m_4$ is chosen. In the bipartite graphs, black/gray edges indicate the chosen/non-chosen assignment. Edge thickness indicates the mean color-concept association rating, but random draws can produce values above/below this mean, resulting in outcomes to left or right of zero. We define \textit{semantic distance} $\Delta S$ as the absolute difference between these probabilities. We have $0 \le \Delta S \le 1$ and a larger $\Delta S$ indicates more certainty in the outcome of the assignment.}
	\label{fig:SDprob}
\end{figure}

We now define semantic distance in the case of two concepts and two colors and illustrate our definition in Fig.~\ref{fig:SDprob} using mango and watermelon as concepts and $m_4$ and $w_4$ as colors.

Given two colors and two concepts, there are two possible assignments of colors to concepts (indicated by black edges on the two bipartite graphs in Fig. \ref{fig:SDprob}). We define the semantic distance to be the absolute difference in the probabilities of each assignment being chosen by a random individual.
Specifically, if $x_1, x_2, x_3, x_4$ are the association ratings between colors and concepts (see Fig.~\ref{fig:SDprob}), we let $\Delta x = (x_{1} + x_{4}) - (x_{2} + x_{3})$. Note that if $\Delta x > 0$, concept M will be assigned with color $m_4$ and concept W will be assigned with color $w_4$. If $\Delta x < 0$, the alternative assignment will be made. We assume each $x_{i}$ is normally distributed, with mean equal to $\bar{x}_{i}$, the mean association across all people for this color and concept, and standard deviation equal to $\sigma_{i} = 1.4 \cdot \bar{x}_{i}(1-\bar{x}_{i})$. This was found to be a good fit to the experimental data\footnote{Many other choices could be made here, by picking other functions that have a similar qualitative shape (i.e., zero standard deviation when $\bar x=0$ or $1$ and maximum standard deviation when $\bar x=0.5$). We experimented with other options and found our results to be robust with respect to the choice of function.}. We define semantic distance as
\begin{equation}\label{deltaS}
\Delta S = \bigl|\, \mathrm{Prob}( \Delta x > 0 ) - \mathrm{Prob}( \Delta x < 0 ) \,\bigr|.
\end{equation}
The probabilities in~\eqref{deltaS} can be computed by computing the $z$-score using the mean and standard deviations described above.
\begin{equation}\label{probDeltax}
\mathrm{Prob}( \Delta x > 0 )
= \Phi\Biggl( \frac{(\bar x_{1} + \bar x_{4}) - (\bar x_{2} + \bar x_{3})}{\sqrt{\sigma_{1}^2+\sigma_{2}^2+\sigma_{3}^2+\sigma_{4}^2}} \Biggr),
\end{equation}
and $\mathrm{Prob}( \Delta x < 0 ) = 1 - \mathrm{Prob}( \Delta x > 0 )$, where $\Phi(\cdot)$ is the cumulative distribution function (cdf) of the standard normal distribution.
The relationship between the $x_i$ and $\Delta x$ is illustrated in Figure~\ref{fig:SDprob}.

Fig.~\ref{fig:Bipartite} left and center illustrate how semantic distance can vary while the assignment problem outcome remains constant. In both examples, concept A is assigned to color 1 and concept B is assigned to color 2, but semantic distance decreases between Fig.~\ref{fig:Bipartite} left and center because the merit distributions overlap more. We predict that such decreases in semantic distance will make visualizations more difficult to interpret. In Fig.~\ref{fig:Bipartite} right, the colors have a greater semantic distance than in Fig.~\ref{fig:Bipartite} center, even though color 1 is more strongly associated than color 2 with both concept A and concept B. Thus, the scenario in Fig.~\ref{fig:Bipartite} right should be more interpretable than Fig.~\ref{fig:Bipartite} center. This example shows how colors can have a large semantic distance with respect to two concepts, even though neither color is strongly evocative of a particular concept within the set. Prior work has shown such cases are easily interpretable \cite{schloss2018} (see Section \ref{sec:asigninf}).

\begin{figure*}[!ht]
	\centering
	\includegraphics[width=1.0\linewidth]{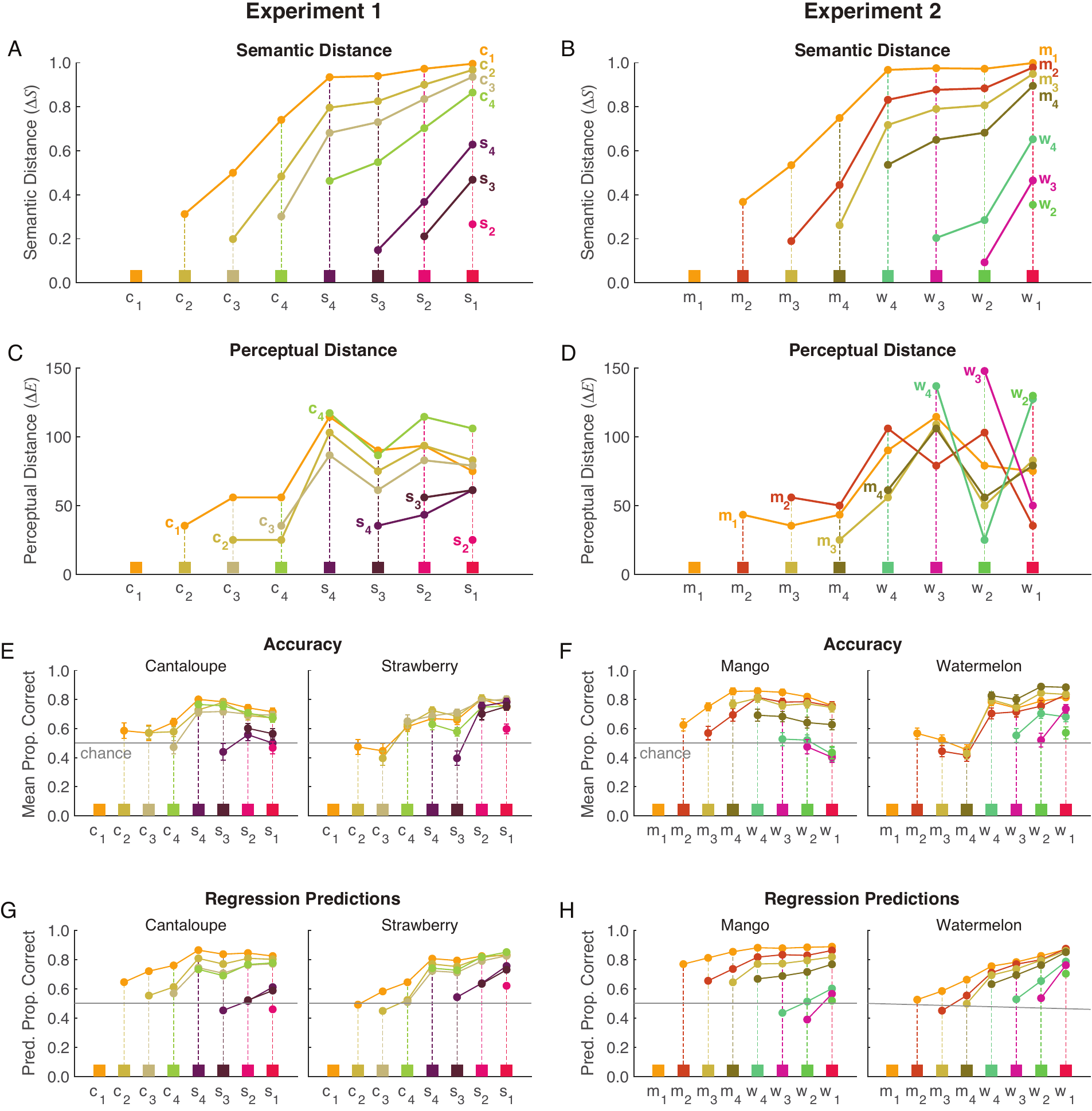}
	\caption{The left column (A,C,E,G) refers to Experiment~1 (cantaloupe and strawberry) and the right column (B,D,F,H) shows the analogous data for Experiment~2 (mango and watermelon). We will describe the left column. The top row shows semantic distances ($\Delta S$) for all pairs of colors. Since $\Delta S$ for a pair of colors is defined in the context of both target fruits, we can represent the data in a single plot. The plot contains 28 points, one for each pair of distinct colors from the set of 8 colors. 
	Each of the 28 points is identified with a pair of colors as follows. The color of the point itself identifies the first color, and the vertical dashed line crossing that point leads to a label on the x-axis, identifying the second color. Thus, all points connected by dashed lines share a common color, and likewise for the solid lines.
	The colored squares along the x-axis and associated labels also serve as a legend for the mark colors.
	In (A), colors $c_{1}$ to $c_{4}$ are the colors most strongly associated with cantaloupe and $s_{1}$ to $s_{4}$ are the colors most associate with strawberry (lower subscripts are more strongly associated with the fruit indicated by the letter, see also Fig.~\ref{fig:FruitColors}). 
	The second row shows perceptual distance ($\Delta E$), plotted in the same manner as semantic distance.
	The third row shows mean proportion of correct responses plotted separately for each target concept because each target was assessed independently. Error bars represent standard errors of the means using the Cousineau \cite{cousineau2005} adjustment to account for overall differences at the subject level.
	The bottom row shows predicted response accuracy using regression equations from Table \ref{table:Acc}.
	\vspace{1cm}}
	\label{fig:PredAcc}
\end{figure*}

Fig.~\ref{fig:PredAcc}A shows semantic distance for all 28 pairwise combinations of 8 colors tested in Experiment 1 (see figure caption for details on how to interpret this plot). There is only one plot for both cantaloupe and strawberry because semantic distance for a given set of colors and concepts is symmetric. That is, the distance between colors $c_{1}$ and $s_{1}$ is the same, regardless of whether the target is cantaloupe or strawberry. When cantaloupe colors are paired with other cantaloupe colors (on curves labeled $c_{1}$, $c_{2}$, $c_{3}$), semantic distance increases as the difference in association strength increases (i.e., distance on the x-axis), but then levels off once reaching strawberry colors because all strawberry colors are similarly weakly associated with cantaloupe. To see the analogous pattern for strawberry colors, it is necessary to compare the heights of data points at each x-axis position. For example, looking at $s_{1}$ on the x-axis, semantic distance steadily increases for pairings with other strawberry colors as association strength difference increases ($s_{2}$ to  $s_{4}$), and then levels off when reaching the four cantaloupe colors.
 
Fig.~\ref{fig:PredAcc}B shows semantic distance for the colors and concepts in Experiment 2. Given that the pattern of association strengths across colors in Experiment 2 (Fig.~\ref{fig:FruitColors}B.1) was similar to Experiment 1 (Fig.~\ref{fig:FruitColors}A.1), the pattern of semantic distances were strongly correlated between the two experiments ($r(26)  = .99, p < .001)$.

\subsubsection{Perceptual discriminability}
We operationalized perceptual discriminability as \textit{perceptual distance} ($\Delta E$) in CIELAB color space, as in previous visualization research \cite{ szafir2018, stone2014}. Fig.~\ref{fig:PredAcc}C shows perceptual distances in Experiment 1 and Fig.~\ref{fig:PredAcc}D shows perceptual distances in Experiment 2, plotted in the same manner as semantic distance. In Experiment 1, perceptual distance (Fig.~\ref{fig:PredAcc}C) deviated from semantic distance (Fig.~\ref{fig:PredAcc}A) but the two variables were still correlated ($r(26)  = .71, p < .001$). In Experiment 2, perceptual distance (Fig.~\ref{fig:PredAcc}D) and semantic distance (Fig.~\ref{fig:PredAcc}B) were uncorrelated ($r(26) = .02, p = .920$). Perceptual distances in Experiment 1 and Experiment 2 were also uncorrelated ($r(26) = .08, p = .673$)

\section{Experiment 1}\label{sec:exp1}
This experiment tested for independent effects of semantic and perceptual distance on interpretability, using the colors in Fig. \ref{fig:FruitColors}A. Although semantic distance and perceptual distance were correlated, we could test for independent effects of each factor using regression analyses.

\subsection{Methods} 

\textbf{Participants.}
36 undergraduates (mean age = 18.3, 25~females, 11~males) participated for credit in Introductory Psychology. All had normal color vision (screened with \cite{hardy2002}), and gave informed consent. The UW--Madison IRB approved the protocol for this study. 

\textbf{Design and Displays.} 
Participants were presented with bar graphs showing fictitious data about preferences for two different fruits (Fig.~\ref{fig:Overview}, left). Each graph had two colored bars, one for each fruit. The bars were two different colors, determined by all 28 pairwise combinations of eight colors (Fig.~\ref{fig:FruitColors}A, Table~\ref{table:ExpColors} in the Supplementary Material). The bars were 50 pixels wide (2.4 cm wide) and varied in height. Each trial contained a taller and shorter bar with base heights of 150 and 100 pixels (5.1 and 3.7 cm), respectively. Bar heights were randomly, and independently, adjusted around their base height by +/- 5 pixels (.2 cm) on each trial. The side of the graph containing the taller bar was left/right balanced. The x and y axes of the graph were 200 and 250 pixels long respectively. The y-axis was labeled as ``Preference'' (font size 14) and the x-axis and bars were unlabeled. The target fruit for a given trial was displayed as text positioned above the graph (20 pt font), centered on the x-axis. Thus, the full experiment design included 2 target concepts (cantaloupe or strawberry) $\times$ 28 color pairs $\times$ 2 positions of the colors within each pair (left or right) $\times$ 2 taller bar sides (left or right) $\times$ 3 repetitions, producing 672 trials.  

The displays were generated and presented using Presentation (\url{www.neurobs.com}). The monitor was a 24.1~in ASUS~ProArt~PA249Q monitor ($1920 \times 1200$ resolution), viewed from about 60~cm. The background was gray (CIE Illuminant D65, x = .3127, y = .3290, Y = $10$~cd/$\text{m}^2$). We used a white point of D65, luminance  = $100$~cd/$\text{m}^2$) to convert between CIELAB and CIE 1931 xyY coordinates. We used a Photo Research PR-655 SpectraScan spectroradiometer to characterize the monitor and verify accurate presentation of the colors. The deviance between the measured and target colors in CIE~1931 xyY coordinates was $< .01$ for x and y, and $< 1$~cd/$\text{m}^2$ for Y.

\textbf{Procedure.} 
 The participants were told that they would be presented with a series of bar graphs, each showing a different person's preferences for two fruits, cantaloupe and strawberry. Within each graph, one bar would represent cantaloupe and the other bar would represent strawberry. The bars would have different colors, but would not be labeled. Above the graph, participants would see the name of one of the two fruits, cantaloupe or strawberry. Their task was to decide which bar corresponded to the fruit described above the graph and to respond by pressing the corresponding arrow key (left or right). Participants were reminded that the bars would not be labeled, and were asked to use their intuition about which bar color corresponded to the fruit described. 

Participants then completed five practice trials drawn at random from all possible trials. They then completed the 672 test trials, presented in a blocked randomized design (three blocks to accommodate three repetitions). Each block included all combinations of targets, color pairs, color positions, and bar height positions, presented in a random order. Participants received a break after each set of 28 trials. Stimuli remained on the screen until participants responded, and trials were separated by a 500-ms inter-trial interval. We recorded which color was chosen and the response time (RT) to make the choice on each trial.

\subsection{Results and Discussion}
We first present results on accuracy, where ``correct'' was defined as the solution to the assignment problem for both possible targets and the two colors on a given trial (see Sections \ref{sec:assign} and \ref{sec:defInterp}). We then present results on RTs, which can be interpreted as how difficult it was to make the decision on each trial, regardless of accuracy.

\textbf{Accuracy.} 
Fig.~\ref{fig:PredAcc}E shows mean accuracy for each color pair when the target was cantaloupe (left) or strawberry (right). To obtain these means, we first calculated the proportion of correct trials for each participant, for each target (cantaloupe or strawberry) and each pair of colors (all 28 combinations of 8 colors). This proportion included 12 data points (2 left/right positions of the colors within each pair $\times$ 2 positions of the taller bar within the bar graph $\times$ 3 repetitions). We then calculated the mean for each target and color pair across participants. 

The correct color for each target and each color pair is indicated by the color positions on the x-axes of Fig. \ref{fig:PredAcc}E. Within each color pair, the color toward the left on the x-axis was correct for cantaloupe, and the color toward the right was correct for strawberry. For example, given $c_{2}$ and $c_{4}$, $c_{2}$ was correct for cantaloupe and $c_{4}$ was correct for strawberry.

We first highlight three key observations in Fig. \ref{fig:PredAcc}E. First, most of the responses were well above chance. This means that participants could reliably decode our encoded mappings, even though there was no legend. Second, there is systematic variability in response accuracy across color pairs. This provides further support that human assignment inference is probabilistic, not deterministic. Recall that if a computer were solving assignment problems in our task, the outcome would be deterministic---responses would all be at 1.0 regardless of whether the assignment problem was robust or fragile (see Sections \ref{sec:assign} and \ref{sec:semantic_distance}). 

Third, the pattern of accuracies resembles aspects of semantic distance (Fig.~\ref{fig:PredAcc}A) and perceptual distance (Fig.~\ref{fig:PredAcc}C). Like both predictors, accuracy tends to be greater when pairs include one cantaloupe color ($c_1$ - $c_4$) and one strawberry color ($s_1$ - $c_4$) (``between-concept pairs''), especially for cantaloupe. The predictors differ in that semantic distance increased monotonically from left to right in \ref{fig:PredAcc}A, whereas perceptual distance is non-monotonic based on how we selected the colors. Perceptual distance is  flatter among ``within-concept'' color pairs where colors are perceptually similar (among $c_1$ - $c_4$ and among $s_1$ - $s_4$), and fluctuates systematically across between-concept color pairs (Fig \ref{fig:PredAcc}C). Accuracy resembles the flatness of perceptual distance for within-concept pairs where colors were most perceptually similar (especially for cantaloupe), but accuracy resembles the smoothness of semantic distance for between-concept pairs where colors were most perceptually distinct. 

To test for independent effects of perceptual and semantic distance on accuracy, we used a mixed effect logistic regression
(R version 4.0.2, lme4 1.1-23). The dependent measure was accuracy on each trial for each participant (1 = correct, 0 = incorrect). We included fixed effects for semantic distance and perceptual distance, and random slopes and intercepts for subjects within each fixed effect. We used z-scores of the predictors in all models to center them and put them on similar scales. As shown in Table~\ref{table:Acc} (Model Acc 1.1), accuracy significantly increased with increased semantic distance and perceptual distance.

Recall that both semantic and perceptual distance are symmetric, they are defined with respect to a given color pair, irrespective of the target. Thus, based on these factors alone, we would predict that the pattern of responses for both targets would be the same. However, as shown in Fig.~\ref{fig:PredAcc}E, there are systematic asymmetries. In particular, note how accuracy among pairs including the strawberry colors was greater for strawberry targets than cantaloupe targets. To fully capture this pattern of data, it is necessary to add another predictor that accounts for differences depending on the target concept.  

Thus, we repeated the same model but added a new factor that could capture target-specific responses: association strength between the target and the correct color. This factor was previously shown to predict accuracy and RTs for similar data \cite{schloss2018}. As shown in Table~\ref{table:Acc} (model Acc 1.2), association strength significantly predicted accuracy, and the previous two factors were still significant. Therefore, accuracy increased with semantic distance, perceptual distance, and association strength between the target and the correct color. Fig.~\ref{fig:PredAcc}G shows the predicted data using weights from model Acc 1.2 in Table~\ref{table:Acc}. The model predictions and data for all 28 color pairs $\times$ 2 concepts are strongly correlated ($r(54) = .82, p < .001$).

We also examined the relation between predictors in the model. Across the 28 color pairs for each of the two targets, association strength between the target and correct color was moderately correlated with semantic distance ($r(54) = .43, p < .001$) and not significantly correlated with perceptual distance ($r(54) = .21, p = .123$).

\begin{table}[tb]
\caption{Mixed-effect logistic regression models of accuracy in Experiment 1 (Acc 1.1, Acc 1.2) and Experiment 2 (Acc 2.1, Acc 2.2).} \label{table:Acc} 
\centering
\begin{tabular}{llrrrr} 
\toprule
\textbf{Model} & \textbf{Factor} & {\textbf{$\beta$}} & \textbf{SE}  & \textit{\textbf{z}} & \textit{\textbf{p}}
\\ \midrule

Acc 1.1 & Intercept      & 0.89 &  0.14  & 6.37 & $<$.001\\
        & PercDist      & 0.22 &  0.06  & 3.58 & $<$.001\\
        & SemDist       & 0.34 &  0.07  & 4.96 & $<$.001 \\
\midrule
Acc 1.2 & Intercept      & 0.91 &  0.14  & 6.38 & $<$.001\\
        & PercDist      & 0.26 &  0.06  & 4.11 & $<$.001\\
        & SemDist       & 0.23 &  0.07  & 3.05 &    .002 \\
        & Assoc         & 0.23 &  0.05  & 4.77 & $<$.001\\ 
\midrule
\midrule
Acc 2.1   & Intercept     & 0.97 &  0.12  & 8.36 & $<$.001\\
        & PercDist      & -0.06 &  0.03  & -1.90 & .057\\
        & SemDist       & 0.55 &  0.06  & 9.24 & $<$.001 \\
\midrule
Acc 2.2  & Intercept     & 1.00  &  0.12  & 8.45 & $<$.001\\
        & PercDist      & -0.06 &  0.03  & -1.84 & .066 \\
        & SemDist       & 0.41  &  0.06  & 6.86 & $<$.001 \\
        & Assoc         & 0.37  &  0.05  & 7.95 & $<$.001\\

\bottomrule 
\end{tabular}
\figskip
\end{table}

\begin{figure*}[ht]
	\centering
	\includegraphics[width=1.0\linewidth]{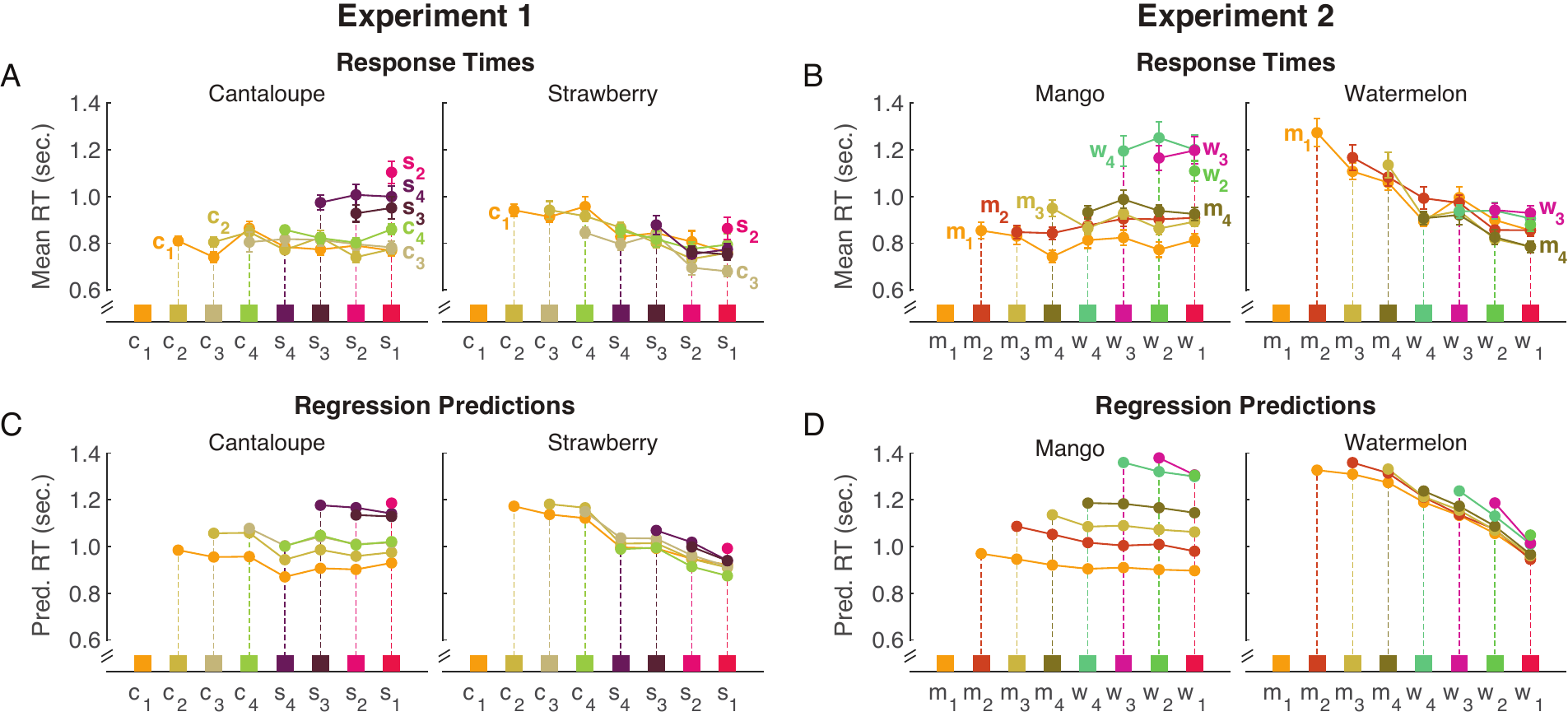}
	\caption{
	The top row shows mean RTs for (A) Experiment 1  and (B) Experiment 2, plotted in the same manner as in Fig. \ref{fig:PredAcc}. Error bars represent standard errors of the means, using the Cousineau \cite{cousineau2005} adjustment to account for overall differences at the subject level. The bottom row shows regression predictions for RTs using the model with all three predictors in Table \ref{table:RT} in (C) Experiment 1 and (D) Experiment 2.}
	\label{fig:RT}
\end{figure*}

\textbf{Response time.} Fig.~\ref{fig:RT}A shows mean RTs for each color pair, obtained by first calculating the median RT across all 12 trials for each target and color pair for each participant, and then calculating the mean over participants. Treating RTs this way avoids effects of outliers without excluding trials \cite{ratcliff1993}. Mean RTs were negatively correlated with mean accuracy ($r(54) = -.71, p < .001$), such that participants responded more quickly for color pairs that facilitate accuracy.

We analyzed the RT data using linear-mixed effect models with the same predictors as for accuracy. The model results are in Table~\ref{table:RT}. When perceptual distance and semantic distance were the only fixed effects (model RT 1.1), neither predictor explained significant variance. When association strength was added to the model, it explained significant variance, as did perceptual distance (model RT 1.2). Thus, RTs were faster when association strength was stronger and when perceptual distance was larger. Fig.~\ref{fig:RT}C shows the model prediction using the weights from model RT 1.2 in Table~\ref{table:RT}. Model predictions were strongly correlated with mean RTs ($r(54) = .82, p < .001$).

\begin{table}[ht]
\caption{Linear mixed-effects regression models of RT in Experiment 1 (RT 1.1, RT 1.2) and Experiment 2 (RT 2.1, RT 2.2).} \label{table:RT} 
\centering
\begin{tabular}{llrrrrr} 
\toprule
\textbf{Model} & \textbf{Factor} & \textbf{$\beta$} & \textbf{SE}  & \textit{\textbf{df}}& \textit{\textbf{t}} & \textit{\textbf{p}}
\\ \midrule

RT 1.1    & Interept      & 1017.7 &  53.2  & 35.0    &   19.1 & $<$.001\\
           & PercDist      & -29.3  &  17.1  & 121.4   &   -1.7 & .088\\
           & SemDist       & -37.2  &  20.6  & 39.0    &   -1.8 & .078\\
         
\midrule

RT 1.2    & Interept      & 1017.7 &  53.2  & 35.0    &   19.1 & $<$.001\\
           & PercDist      & -42.8  &  17.0  & 156.9   &   -2.5  & .013\\
           & SemDist       & 1.9    &  19.6  & 78.6    &    0.1 & .924\\
           & Assoc         & -68.3  &  15.9  & 48.0    &   -4.3 & $<$.001\\
\midrule
\midrule
RT 2.1     & Intercept     & 1121.5 &  62.3  & 35.0 & 18.0 &   $<$.001\\
            & PercDist      & 12.1   &  8.0   & 35.0 & 1.5  &   .139\\
            & SemDist       & -86.4  &  15.1  & 35.0 & -5.7 &   $<$.001 \\
\midrule            
RT 2.2     & Intercept     & 1121.5 &  62.3  & 35.0 & 18.0 &   $<$.001\\
            & PercDist      & 9.0    &  7.9   & 35.0 & 1.1  &   .260\\
            & SemDist       & -36.0  &  9.9   & 35.0 & -3.6 &   $<$.001 \\
            & Assoc         & -120.6 &  20.0  & 35.0 & -6.0 &   $<$.001 \\

\bottomrule 
\end{tabular}
\figskip
\end{table}

In summary, semantic distance and perceptual distance both independently contributed to interpretability. However, the ``cantaloupe colors'' and ``strawberry colors'' were clustered in different parts of color space, (Fig.~\ref{fig:FruitColors}A.2-3), so conflicts between semantic and perceptual distance were only minor. In Experiment 2, we address the question of how these two factors would contribute to interpretability if they were overall decorrelated and included examples of large conflicts. 


\section{Experiment 2}\label{sec:exp2}

This experiment tested for independent effects of semantic and perceptual distance when these two factors were uncorrelated. The pattern of association strengths was similar to Experiment 1 (Fig.~\ref{fig:FruitColors}A.1 and B.1), so the pattern of semantic distances were also similar (Fig.~\ref{fig:PredAcc}A and \ref{fig:PredAcc}B). However, the relative locus of colors in CIELAB space was different. In Experiment 1, the strong associates for each concept clustered together (Fig.~\ref{fig:FruitColors}A.2-A.3), but in Experiment 2, the four ``watermelon colors'' were split on either side of the ``mango colors'' along the a* axis (Fig.~\ref{fig:FruitColors}B.2-B.3). Thus, for watermelon, the most semantically similar colors were furthest in color distance.

\subsection{Methods}
36 undergraduates (mean age = 19.4, 18~females, 18~males) participated for credit in Introductory Psychology. All had normal color vision (screened with \cite{hardy2002}), and gave informed consent. The design, displays, and procedure were the same as Experiment 1, except we tested the colors and fruits in Fig.~\ref{fig:FruitColors}B (Table~\ref{table:ExpColors}). 

\subsection{Results and discussion}
The colors and fruits in Experiment 1 and 2 differed in that their patterns of perceptual distances were uncorrelated ($r(26) = .08, p = .673$) but their patterns of semantic distances were almost perfectly correlated ($r(26)  = .99, p < .001)$. Thus, if the patterns of data in Experiment 2 are similar to Experiment 1, they can be attributed to their similarities in semantic distance. Fig.~\ref{fig:PredAcc}F shows the mean accuracy data and Fig.~\ref{fig:RT}B shows the mean RTs, calculated in the same manner as in Experiment~1. Indeed, there were significant correlations between the patterns of accuracy ($r(54) = .66, p < .001$) and RT ($r(54) = .79, p < .001$) between the two experiments. 

\textbf{Accuracy.} We analyzed accuracy using the same mixed-effect logistic regression models as in Experiment 1. The first model including perceptual distance and semantic distance showed that semantic distance significantly predicted accuracy (Table~\ref{table:Acc}, model Acc 2.1). The effect of perceptual distance was marginal, but it was in the opposite direction from Experiment 1. That is, accuracy tended to increase for more perceptually \textit{similar} colors, probably because colors that were perceptually similar (e.g., $w_{1}$ (red) and $m_{2}$ (dark orange)) were semantically different, whereas colors that were perceptually distant were semantically similar (e.g., $w_{1}$ (red) and $w_{2}$ (green)) (Fig.~\ref{fig:Overview}).

When we added association strength between the target and correct color into the model, association strength was a significant predictor, as was semantic distance (Table~\ref{table:Acc}, model Acc 2.2). The effect of perceptual distance was still marginal, again in the opposite direction (more perceptually different tended to result in reduced accuracy). Fig.~\ref{fig:PredAcc}H shows the predicted accuracy based on the regression weights in model Acc 2.2. The model predictions strongly correlated with the mean accuracy data in Fig.~\ref{fig:PredAcc}F ($r(54) = .83, p <.001).$ In this experiment, association strength between the target and correct color was again moderately correlated with semantic distance ($r(54) = .42, p = .001$) and not significantly correlated with perceptual distance ($r(54) = -.02, p = .900$).

\textbf{Response time.} As in Experiment 1, RT and accuracy were negatively correlated ($r(54) = -.82, p < .001$), indicating it was easier to make decisions for color pairs that facilitated accuracy. We analyzed RTs using the same linear mixed-effect models from Experiment 1. The first model including only perceptual distance and semantic distance showed a significant effect of semantic distance and no effect of perceptual distance (Table \ref{table:RT}, model RT 2.1). Adding association strength between the target and the correct color (model RT 2.2) resulted in significant effects of association strength and semantic distance but still not perceptual distance. Fig.~\ref{fig:RT}D shows the predicted RTs based on the regression weights in model RT 2.2. The model predictions strongly correlated with the mean RTs in Fig.~\ref{fig:RT}B ($r(54) = .88, p <.001).$

In summary, semantic distance dominated interpretability when these two factors were uncorrelated overall. Perceptual distance had a marginal effect, but it was in the opposite direction from what might be expected (i.e., smaller perceptual distances tended to be more interpretable). This was because the stimulus set included cases with strong conflicts, such that large semantic distances amounted to small perceptual distances (especially for watermelon), and under such conflicts greater semantic distance resulted in greater interpretability. 

\section{General Discussion and Conclusion}
In this study we tested whether people's ability to interpret color palettes in information visualizations depended on semantic distance, independent of perceptual distance. The results of both experiments demonstrated that increasing semantic distance improved interpretability, independent of variation in perceptual distance. In Experiment 1, we selected colors such that perceptual and semantic distance co-varied: the four colors that were most strongly associated with cantaloupe were clustered separately from the colors most strongly associated with strawberry. Under these conditions, both semantic and perceptual distance independently contributed to increased interpretability. In Experiment 2, we selected the colors in a way that decoupled perceptual and semantic distance: the four colors that were most strongly associated with mango were between the colors most strongly associated with watermelon on the a* plane of CIELAB space. Across all color pairs in Experiment 2, perceptual distance and semantic distance were uncorrelated, but there were cases in which these two factors were in direct conflict (Fig. \ref{fig:Overview}). In this experiment, accuracy and RT both improved with increased semantic distance, with no significant effects of perceptual distance. The results of this study suggest that it may be worth relaxing constraints on perceptual distance in favor of maximizing semantic distance to create interpretable color palettes.
 
We studied colors that were distant enough ($\Delta E \geq 25$) to be noticeably different \cite{stone2014, szafir2018}, but we expected that perceptual distance would play a larger role if distances were smaller. However, if colors were no longer perceptually discriminable, they would also no longer be semantically discriminable.  Thus, thresholding at some degree of semantic distance may be sufficient to ensure both perceptual and semantic discriminability. Certainly, there is some lower threshold at which perceptual and semantic discriminability would be too small for interpretability, but there also may be an upper threshold at which further increasing perceptual or semantic discriminability would have no further benefit. Substantial work has investigated lower thresholds for perceptual discriminability for information visualizations \cite{stone2014, szafir2018}, but future work is needed to understand thresholds for semantic discriminability. Moreover, as in prior visualization work \cite{stone2014, szafir2018} we used $\Delta E$ as our perceptual distance metric, but future work could evaluate whether different perceptual distance metrics (e.g., CIEDE2000) are better at predicting interpretability. 

As part of this study, we developed semantic distance, $\Delta S$, as a metric to quantify semantic discriminability between pairs of colors and concepts. Semantic distance is the absolute difference in the probabilities that a random observer will make each of the two possible assignments, where the randomness is due to inherent variability in association strengths across individuals. Quantifying semantic discriminability becomes more difficult when there are more than two colors or two concepts because there are more than two possible assignments. Solving assignment problems becomes more complicated in this case, and we cannot write a simple formula as in~\eqref{probDeltax} to compute assignment probabilities.\footnote{Generally, solving a large assignment problem (many colors and concepts) cannot be reduced to solving a sequence of smaller two-color two-concept assignment problems. For example, the colors (red,yellow) and concepts (apple,banana) have a straightforward assignment, but if we add another color and concept: (red,yellow,green) and (apple,banana,strawberry), then apple switches from red to green due to the presence of strawberry.} 
Possible approaches for quantifying semantic discriminability for more than two colors and concepts could involve obtaining a distribution over possible assignments (e.g., via Monte Carlo simulation), resulting from uncertainty in color-concept association ratings and applying one of many possible metrics. For example, if we used \emph{entropy}, maximum entropy would correspond to the fragile case of all assignments having equal probability. Conversely, minimum entropy would correspond to the robust case of one assignment having a very high probability and all other assignments having near-zero probability.

In this study, we used mean human color-concept association ratings to quantify association strengths. However, efficient automated approaches exist for estimating color-concept associations using images \cite{lindner2012a, lindner2012b, lin2013, rathore2020} and natural language databases \cite{setlur2016, havasi2010}. Different methods can be used to extract colors from images, but evidence suggests methods that leverage perceptual dimensions of color and cognitive representations of color categories are best for estimating human color-concept associations
\cite{rathore2020}. Such estimates, combined with an appropriate method for quantifying variance in the sample images, could be used as input to  calculate semantic distance.

The results of this study can help with designing interpretable color palettes, but interpretability is only one of the many goals in color palette design. Other priorities might include helping observers (1) locate a target in visual search \cite{treisman1980, healey1996, wolfe2017, haroz2012, gramazio2014}, (2) estimate the area of colored regions \cite{albers2014, gramazio2017}, (3) refer to the colors easily by name \cite{heer2012}, (4) appreciate the visualization aesthetically \cite{gramazio2017}, or (5) obtain an affective impression from the overall palette \cite{bartram2017}. Different design properties are relevant for these different priorities. For example, the ability to estimate the relative area occupied by colored regions increases with perceptual distance between colors, but aesthetic preferences for those same visualizations decreases with perceptual distance \cite{gramazio2017}. Extensive work is needed to understand how to navigate such trade-offs in palette design, depending on the priorities and format of a given visualization \cite{gramazio2017, bartram2017, stone2014, lin2013}. The present work provides a step in that direction by showing that maximizing perceptual distance is not necessary for creating interpretable color palettes, leaving room for maximizing the other factors that contribute to effective palette design.

\acknowledgments{
We thank Ragini Rathore, Kevin Lande, Kushin Mukherjee, Melissa Schoenlein, Anna Bartel, Brian Yin, and Joris Roos for their feedback on this work, and Shannon Sibrel, Autumn Wickman, Marin Murack, Yuke Liang, Charles Goldring, and Brianne Sherman for help with data collection. This work was supported by the UW--Madison Office of the Vice Chancellor for Research and Graduate Education and Wisconsin Alumni Research Foundation.}


\clearpage

\bibliographystyle{abbrv}

\bibliography{VIS2020_SemDiscrim.bib}

\clearpage
\appendix
\renewcommand*{\thesection}{S}
\counterwithin{figure}{section}
\counterwithin{table}{section}
\section{Supplementary Material}\label{sec:supplementary}

\subsection{Isolated and balanced palettes in Schloss et al.~\cite{schloss2018}}

Schloss et al.~\cite{schloss2018} used a recycling task to study how people might estimate merit during assignment inference. By creating palettes according to different merit functions and determining which was easier for participants to interpret, it was possible to infer properties about the type of merit people use in assignment inference. Each palette had six colors for each of six types of objects to be discarded (paper, plastic, glass, metal, compost, and trash). 

There were two different merit functions. The first, \textit{isolated} merit function, simply uses the association strength as the merit function. This choice does not consider association strengths between unassigned color and concept pairs. For example, this would assign plastic to light gray and glass to light blue, even though plastic was also strongly associated with light blue and glass was also strongly associated with light gray. The second, \textit{balanced} merit function, uses the difference between the association strength of the assigned color-concept pair and the largest association strength among all pairs involving the same concept but every other available color. For example, this would assign plastic to red and glass to cyan because plastic was weakly associated with cyan and glass was weakly associated with red. Using our present terminology, the color palettes generated using the balanced merit function would be more semantically discriminable because the difference score in the computation minimizes confusability between assigned and unassigned color pairs. We note that these two merit functions make distinct assignments for assignment problems involving more than two colors or concepts, but in the case of two colors and concepts, as in the present study, both merit functions produce identical assignments.

A key observation in \cite{schloss2018} was that simply using association strengths as the merit scores was inadequate because that did not account for the fact that the same color may be highly associated with several concepts. The results suggested that assignment inference benefited more from semantic discriminability than from maximizing association strength, given that accuracy was greater for the balanced merit function than the isolated merit function.  

\begin{figure}[ht!!]
	\centering
	\includegraphics[width=1.0\columnwidth]{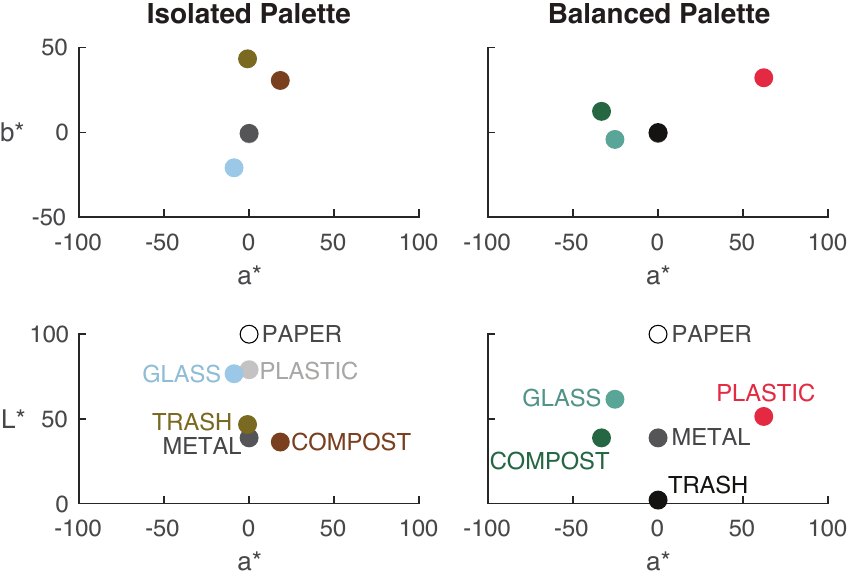}
	\caption{The six colors in the isolated palette and balanced palette from Schloss et al.~\cite{schloss2018} (Experiment 2), plotted in CIELAB space on the a*, b* plane (top row) and a*, L* plane (bottom row). Not all colors are visible on the a*, b* plane (i.e., overlapping a*, b* coordinates). Labels indicate the ``correct'' assignment within each palette, determined by solving the encoding assignment problem according to each merit function.}
	\label{fig:RecColors}
\end{figure}

However, there was a limitation to this prior work, which we address in the present study. Fig.~\ref{fig:RecColors} shows the colors for the isolated palette and balanced palette in CIELAB space.  The colors in the isolated color palette were close together in this perceptual color space, whereas colors in the balanced color palette were more distant. Thus, it is unclear if colors in the balanced color palette were easier to interpret because they were more semantically discriminable, or because they were more perceptually discriminable.

\subsection{Coordinates for colors tested in the present study}
Table \ref{table:ExpColors} shows the CIE 1931 xyY and CIELAB coordinates for each of the colors tested in the present study. 

\begin{table}[ht!!]
\caption{CIE 1931 xyY and CIELAB coordinates for colors in this study [white point CIE Illuminant D65 (x = 0.313, y = 0.329, Y = 100)]. Labels in ``Color'' use the notation in Fig.~\ref{fig:FruitColors} for cantaloupe ($c$), strawberry ($s$), mango ($m$), and watermelon ($w$). Numbers in \# are UW-58 color notation. These colors are a subset of the colors in Table \ref{table:UW_58_colors}} 
\label{table:ExpColors}
\begin{tabular}
{cr S[table-format=1.3]
    S[table-format=1.3]
    S[table-format=2.2]
    S[table-format=2.0]
    S[table-format=3.2]
    S[table-format=3.2]}
\toprule
\multicolumn{1}{c}{\textbf{Color}} &
\multicolumn{1}{c}{\textbf{\#}} &
\multicolumn{1}{c}{\textbf{x}} & \multicolumn{1}{c}{\textbf{y}} & \multicolumn{1}{c}{\textbf{Y}} & \multicolumn{1}{c}{\textbf{L*}} & \multicolumn{1}{c}{\textbf{a*}} & \multicolumn{1}{c}{\textbf{b*}}\\
\midrule
$c_{1}$  &  58 & 0.492    & 0.443    & 48.28     & 75   & 21.04     & 76.21    \\ 
$c_{2}$  &  50 & 0.409    & 0.439    & 48.28     & 75   & -2.62     & 49.93    \\ 
$c_{3}$  &  39 & 0.364    & 0.386    & 48.28     & 75   & -1.31     & 24.97    \\ 
$c_{4}$  &  46 & 0.360    & 0.471    & 48.28     & 75   & -27.58    & 48.62    \\ 
$s_{4}$  &  8 & 0.369     & 0.181    & 4.42      & 25   & 51.24     & -22.35    \\ 
$s_{3}$  &  28 & 0.410    & 0.291    & 4.42      & 25   & 24.97     & 1.31    \\ 
$s_{2}$  &  32 & 0.492    & 0.257    & 18.42     & 50   & 74.90     & 3.93    \\ 
$s_{1}$  &  44 & 0.567    & 0.299    & 18.42     & 50   & 73.59     & 28.89    \\ 
\midrule 
$m_{1}$  &  58 & 0.492    & 0.443    & 48.28     & 75   & 21.04     & 76.21     \\ 
$m_{2}$  &  53 & 0.563    & 0.373    & 18.42     & 50   & 47.32     & 52.55    \\ 
$m_{3}$  &  50 & 0.409    & 0.439    & 48.28     & 75   & -2.62     & 49.93    \\ 
$m_{4}$  &  49 & 0.437    & 0.472    & 18.42     & 50   & -2.62     & 49.93    \\ 
$w_{4}$  &  36 & 0.270    & 0.433    & 48.28     & 75   & -51.24    & 22.35    \\ 
$w_{3}$  &  10 & 0.408    & 0.211    & 18.42     & 50   & 76.21     & -21.04  \\ 
$w_{2}$  &  48 & 0.310    & 0.503    & 48.28     & 75   & -52.55    & 47.32    \\ 
$w_{1}$  &  44 & 0.566    & 0.299    & 18.42     & 50   & 73.59     & 28.89    \\ 

\bottomrule

\end{tabular}
\end{table}

\subsection{Color-concept association data}
\begin{figure}[ht!]
	\centering 
	\includegraphics[width=1.0\columnwidth]{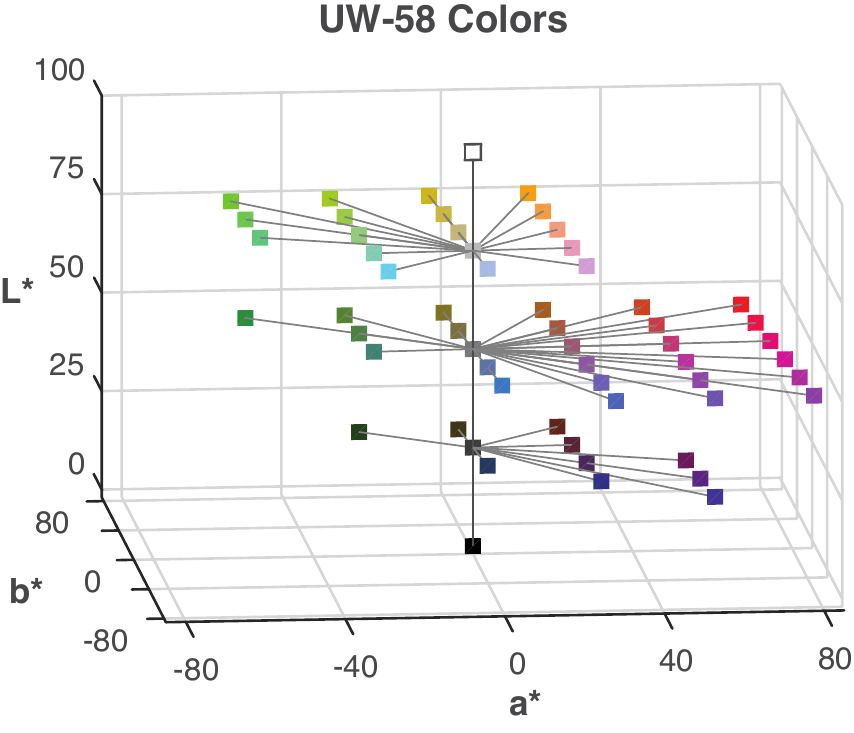}
	\caption{UW-58 colors plotted in CIELAB color space (figure from \cite{rathore2020}). See Table~\ref{table:UW_58_colors} for the CIELAB coordinates.}
	\label{fig:UW58}
\end{figure}

The color-concept association dataset used to select the colors and generate the model predictors in the present experiment was from Rathore et al.~\cite{rathore2020}. The dataset includes 54 participants' ratings of how  much they associated each of the UW-58 colors with each of 12 fruit concepts. The UW-58 colors were defined by superimposing a 3D grid in CIELAB space with a distance $\Delta = 25$ in each dimension (L*, a*, and b*), and then selecting all points whose colors could be rendered in RGB space (using MATLAB's \texttt{lab2rgb} function). The grid was rotated around the achromatic axis to get as many RGB valid high-chroma colors as possible (3\degree\ rotation included the largest set). Fig. \ref{fig:UW58} shows the UW-58 colors plotted in CIELAB space and Table \ref{table:UW_58_colors} contains the coordinates in CIE 1931 xyY, CIELAB (L*,a*,b*) and CIELCh (L*,C*,h) color spaces. The background was gray (CIE Illuminant D65, x = .3127, y = .3290, Y = $10$~cd/$\text{m}^2$).

For each concept, participants saw each color one at a time in a random order. The concept name was presented at the top of the screen and the colored square ($100 \times 100$ pixels) was centered below. Participants rated each color twice for a given concept before going on to the next concept in a blocked randomized design (all colors were rated once before being repeated). The order of the concepts was randomized across participants. To make their ratings, participants used a continuous line-mark slider scale displayed below the colored square. The left end point was labeled ``not at all'', the right end point was labeled ``very much'', and the midpoint was marked with a vertical divider.  Participants responded by sliding their cursor along the response scale and clicking to record their response. There was a 500~ms blank gray screen between each trial. Prior to starting the experiment, participants were instructed to ``anchor'' the end points of the color scale for each concept. This was done so they knew what associations of ``not at all'' and ``very much'' meant for them in the context of these concepts and colors \cite{palmer2013}. During anchoring, participants saw a display of all UW-58 colors and a list of concepts. The experimenter instructed them to  point to the color they associated most/least with each concept, and told them to rate those colors near the ``not at all''/``very much'' endpoints of the scale (respectively) during the experiment. The monitor, viewing conditions, and calibration were the same as reported in the main text of the present study. 

\begin{figure*}[ht!]
	\centering 
	\includegraphics[width=1.0\linewidth]{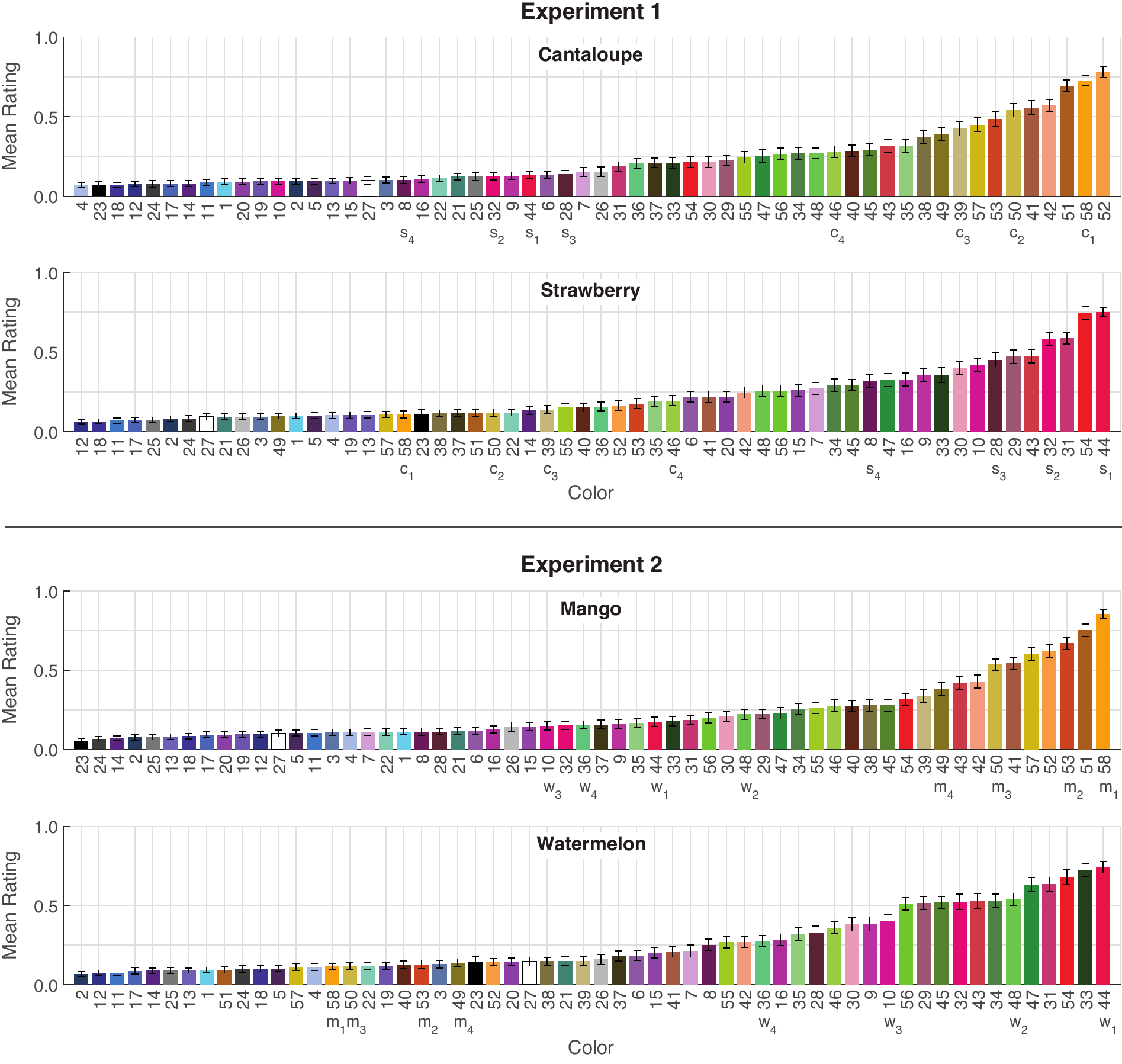}
	\figskip\caption{Mean color-concept associations for each of the UW-58 colors with each of the concepts from Experiment~1 (cantaloupe and strawberry) and Experiment~2 (mango and watermelon). Data are sorted from low to high for each concept. The number below each bar refers to the color's number in the UW-58 color set (see Table \ref{table:UW_58_colors} for color coordinates and Table \ref{table:assocData} for means used to create this figure). Labels below the numbers on the x-axis indicate colors that were tested in the present study (see Fig. \ref{fig:FruitColors} in the main text and Table \ref{table:ExpColors}). Error bars represent standard errors of the means. These data are from  \cite{rathore2020}.}
	\figskip\label{fig:fruitAssoc}
\end{figure*}

Supplementary Fig.~\ref{fig:fruitAssoc} shows the mean color-concept associations for the two fruits used in Experiment 1 (cantaloupe and strawberry) and the Experiment 2 (mango and watermelon). Supplementary Table~\ref{table:assocData} includes the mean association ratings for these four fruits, plus the other eight fruits not used in the present study. The full dataset including individual participant ratings can be found at:\\ \url{github.com/SchlossVRL/semantic-discriminability}.

\subsection{Models with association strength only}
In the main text we report on two sets of mixed-effect models of accuracy (Table \ref{table:Acc}) and RT (Table \ref{table:RT}), one set using perceptual and semantic distance as predictors, and another set including those two predictors, plus association strength between the correct color and the target (hereforth, ``association strength''). Here, we report on models that only include association strength. Table \ref{table:Acc_Assoc} shows that accuracy significantly increased as association strength increased in Experiment 1 (Acc 1.A) and Experiment 2 (Acc 2.A). Likewise, Table \ref{table:RT_Assoc} shows that RTs significantly decreased as association strength increased in Experiment 1 (RT 1.A) and Experiment 2 (RT 2.A). These results are consistent with previous evidence that participants are more accurate and faster at decoding encoded mappings with increased association strength between the target and the correct color \cite{schloss2018}.

\begin{table}[ht]
\caption{Mixed-effect logistic regression models of accuracy using only association strength (A) as a predictor in Experiment 1 (Acc 1.A) and Experiment 2 (Acc 2.A).} \label{table:Acc_Assoc} 
\centering
\begin{tabular}{llrrrr} 
\toprule
\textbf{Model} & \textbf{Factor} & {\textbf{$\beta$}} & \textbf{SE}  & \textit{\textbf{z}} & \textit{\textbf{p}}
\\ \midrule

Acc 1.A & Intercept     & 0.79 &  0.12  & 6.73 & $<$.001\\
        & Assoc         & 0.37 &  0.05  & 6.84 & $<$.001\\ 
\midrule
\midrule
Acc 2.A  & Intercept    & 0.93  &  0.10  & 8.98 & $<$.001\\
         & Assoc        & 0.53  &  0.05  & 10.67 & $<$.001\\

\bottomrule 
\end{tabular}
\figskip
\end{table}

\begin{table}[ht]
\caption{Linear mixed-effects regression models of RT using only association strength (A) as a predictor in Experiment 1 (RT 1.A) and Experiment 2 (RT 2.A).} \label{table:RT_Assoc} 
\centering
\begin{tabular}{llrrrrr} 
\toprule
\textbf{Model} & \textbf{Factor} & \textbf{$\beta$} & \textbf{SE}  & \textit{\textbf{df}}& \textit{\textbf{t}} & \textit{\textbf{p}}
\\ \midrule

RT 1.A      & Intercept      & 1017.7 &  53.2  & 35.0    &   19.1 & $<$.001\\
            & Assoc.         & -76.4  &  17.0  & 35.0    &   -4.5 & $<$.001\\
\midrule
\midrule
RT 2.A      & Intercept     & 1121.5 &  62.3  & 35.0 &  18.0   &   $<$.001\\
            & Assoc.        & -135.7 &  21.8  & 35.0 &  -6.2   &   $<$.001\\

\bottomrule 
\end{tabular}
\figskip
\end{table}

Using the intercept and weights in Table \ref{table:Acc_Assoc} and Table \ref{table:RT_Assoc}, we predicted accuracy and RT based on association strength alone, and correlated these predictions with the accuracy and RT data in Fig. \ref{fig:PredAcc} and Fig. \ref{fig:RT}, respectively. These correlations are shown in Table \ref{table:fits}. We also repeated this procedure using models that included perceptual and semantic distance only (Acc 1.1, Acc 2.1, RT 1.1, RT 2.1) from the main text. In most cases, models with only association strength fit the data better than models with perceptual and semantic distance, even though models with only association strength had fewer parameters. 

However, these results should not be taken to mean that association strength is generally more important that perceptual and semantic distance for predicting accuracy and RT. Given the design of the present experiments, the color that was most strongly associated with the target was almost always the correct answer, so association strength was a reliable cue to optimal assignment. Yet, cases can arise in which local association strengths conflict with global assignments (see Section \ref{sec:asigninf}), in which association strengths would be a poor cue for optimal assignment. To fully understand the effects of association strength and semantic distance, it would be necessary to vary them orthogonally (i.e., make association strength a valid cue to optimal assignment on only half of the trials, and independently vary cue validity with semantic distance).

\begin{table}[ht]
\caption{Correlations between model predictions and data for models using association strength as the only predictor  or perceptual and semantic distances as the predictors. For all correlations, $df = 54, p < .001$. } \label{table:fits} 
\centering
\begin{tabular}{lcc} 
\toprule

Dataset &  Assoc.  & PercDist.+SemDist.  \\
 
\midrule

Acc Exp. 1  &   .59   &   .74   \\
Acc Exp. 2  &   .71   &   .69   \\
\midrule
RT Exp. 1  &   .78   &   .52    \\
RT Exp. 2  &   .85   &   .55    \\

\bottomrule 
\end{tabular}
\figskip
\end{table}

\begin{table*}[tb]
\sisetup{round-mode=places}
\centering
\figskip\caption{Coordinates for the University of Wisconsin 58 (UW-58) colors in CIE 1931 xyY space, CIELAB color space, and CIELch color space. The ``L*'' in CIELAB and CIELch is the same. The white point used to convert between CIE 1931 xyY and CIELAB space was CIE Illuminant D65 (x = 0.313, y = 0.329, Y = 100). Table reproduced from Rathore et al.~\cite{rathore2020}.}
\label{table:UW_58_colors}
\begin{tabular}
{c*{1}{
    S[round-precision=3]
    S[round-precision=3]
    S[round-precision=2]
    S[round-precision=2]
    S[round-precision=2]
    S[round-precision=2]
    S[round-precision=2]
    S[round-precision=2]
}}
\toprule
\multicolumn{1}{c}{\textbf{Color}} &
\multicolumn{1}{c}{\textbf{x}} & \multicolumn{1}{c}{\textbf{y}} & \multicolumn{1}{c}{\textbf{Y}} & \multicolumn{1}{c}{\textbf{L*}} & \multicolumn{1}{c}{\textbf{a*}} & \multicolumn{1}{c}{\textbf{b*}} & \multicolumn{1}{c}{\textbf{C*}} & \multicolumn{1}{c}{\textbf{h}} \\
\midrule
1 & 0.22397    & 0.28399    & 48.278     & 75.         & -23.657    & -26.274    & 35.355     & 228.        \\ 
2 & 0.2081     & 0.21415    & 4.4155     & 25.         & 1.3084     & -24.966    & 25.00         & 273.        \\ 
3 & 0.24471    & 0.25394    & 18.419     & 50.         & 1.3084     & -24.966    & 25.00         & 273.        \\ 
4 & 0.26261    & 0.27354    & 48.278     & 75.         & 1.3084     & -24.966    & 25.00         & 273.        \\ 
5 & 0.28644    & 0.19884    & 4.4155     & 25.         & 26.274     & -23.657    & 35.355     & 318.        \\ 
6 & 0.29797    & 0.24051    & 18.419     & 50.         & 26.274     & -23.657    & 35.355     & 318.        \\ 
7 & 0.30272    & 0.2622     & 48.278     & 75.         & 26.274     & -23.657    & 35.355     & 318.        \\ 
8 & 0.36941    & 0.18108    & 4.4155     & 25.         & 51.24      & -22.349    & 55.902     & 336.43     \\ 
9 & 0.35288    & 0.2259     & 18.419     & 50.         & 51.24      & -22.349    & 55.902     & 336.43     \\ 
10 & 0.40795    & 0.21059    & 18.419     & 50.         & 76.206     & -21.041    & 79.057     & 344.56     \\ 
11 & 0.18715    & 0.19157    & 18.419     & 50.         & 2.6168     & -49.931    & 50.00      & 273.        \\ 
12 & 0.19063    & 0.1298     & 4.4155     & 25.         & 27.583     & -48.623    & 55.902     & 299.57     \\ 
13 & 0.23146    & 0.18448    & 18.419     & 50.         & 27.583     & -48.623    & 55.902     & 299.57     \\ 
14 & 0.25495    & 0.12284    & 4.4155     & 25.         & 52.548     & -47.315    & 70.711     & 318.        \\ 
15 & 0.27871    & 0.17635    & 18.419     & 50.         & 52.548     & -47.315    & 70.711     & 318.        \\ 
16 & 0.32783    & 0.1674     & 18.419     & 50.         & 77.514     & -46.006    & 90.139     & 329.31     \\ 
17 & 0.17813    & 0.14021    & 18.419     & 50.         & 28.891     & -73.589    & 79.057     & 291.43     \\ 
18 & 0.1742     & 0.082515   & 4.4155     & 25.         & 53.857     & -72.28     & 90.139     & 306.69     \\ 
19 & 0.21726    & 0.13588    & 18.419     & 50.         & 53.857     & -72.28     & 90.139     & 306.69     \\ 
20 & 0.25909    & 0.13088    & 18.419     & 50.         & 78.822     & -70.972    & 106.07     & 318.        \\ 
21 & 0.25332    & 0.35108    & 18.419     & 50.         & -24.966    & -1.3084    & 25.00      & 183.        \\ 
22 & 0.26938    & 0.34523    & 48.278     & 75.         & -24.966    & -1.3084    & 25.00      & 183.        \\ 
23 & 0.31273    & 0.32902    & 0.00       & 0.          & 0.00       & 0.00          & 0.00       & 0.00          \\ 
24 & 0.31273    & 0.32902    & 4.4155     & 25.         & 0.00       & 0.00          & 0.00    & 0.00          \\ 
25 & 0.31273    & 0.32902    & 18.419     & 50.         & 0.00       & 0.00          & 0.00    & 0.00          \\ 
26 & 0.31273    & 0.32902    & 48.278     & 75.         & 0.00       & 0.00          & 0.00    & 0.00          \\ 
27 & 0.31273    & 0.32902    & 100.00     & 100.        & 0.00       & 0.00          & 0.00          & 0.00          \\ 
28 & 0.41044    & 0.2905     & 4.4155     & 25.         & 24.966     & 1.3084     & 25.        & 3.00          \\ 
29 & 0.37353    & 0.30534    & 18.419     & 50.         & 24.966     & 1.3084     & 25.        & 3.00          \\ 
30 & 0.3568     & 0.31196    & 48.278     & 75.         & 24.966     & 1.3084     & 25.        & 3.00          \\ 
31 & 0.43376    & 0.28095    & 18.419     & 50.         & 49.931     & 2.6168     & 50.        & 3.00          \\ 
32 & 0.49181    & 0.25666    & 18.419     & 50.         & 74.897     & 3.9252     & 75.        & 3.00          \\ 
33 & 0.3075     & 0.52435    & 4.4155     & 25.         & -26.274    & 23.657     & 35.355     & 138.00        \\ 
34 & 0.31561    & 0.44408    & 18.419     & 50.         & -26.274    & 23.657     & 35.355     & 138.00        \\ 
35 & 0.31654    & 0.41004    & 48.278     & 75.         & -26.274    & 23.657     & 35.355     & 138.00        \\ 
36 & 0.27022    & 0.43268    & 48.278     & 75.         & -51.24     & 22.349     & 55.902     & 156.43     \\ 
37 & 0.41792    & 0.44961    & 4.4155     & 25.         & -1.3084    & 24.966     & 25.        & 93.00         \\ 
38 & 0.38179    & 0.40728    & 18.419     & 50.         & -1.3084    & 24.966     & 25.        & 93.00         \\ 
39 & 0.36355    & 0.38634    & 48.278     & 75.         & -1.3084    & 24.966     & 25.        & 93.00         \\ 
40 & 0.52174    & 0.37656    & 4.4155     & 25.         & 23.657     & 26.274     & 35.355     & 48.00        \\ 
41 & 0.44682    & 0.36993    & 18.419     & 50.         & 23.657     & 26.274     & 35.355     & 48.00         \\ 
42 & 0.41032    & 0.36214    & 48.278     & 75.         & 23.657     & 26.274     & 35.355     & 48.00         \\ 
43 & 0.50874    & 0.33341    & 18.419     & 50.         & 48.623     & 27.583     & 55.902     & 29.566     \\ 
44 & 0.56618    & 0.29873    & 18.419     & 50.         & 73.589     & 28.891     & 79.057     & 21.435     \\ 
45 & 0.36753    & 0.52508    & 18.419     & 50.         & -27.583    & 48.623     & 55.902     & 119.57     \\ 
46 & 0.35998    & 0.47135    & 48.278     & 75.         & -27.583    & 48.623     & 55.902     & 119.57     \\ 
47 & 0.29736    & 0.57731    & 18.419     & 50.         & -52.548    & 47.315     & 70.711     & 138.00        \\ 
48 & 0.31049    & 0.50294    & 48.278     & 75.         & -52.548    & 47.315     & 70.711     & 138.00        \\ 
49 & 0.43671    & 0.47238    & 18.419     & 50.         & -2.6168    & 49.931     & 50.         & 93.00         \\ 
50 & 0.4092     & 0.43925    & 48.278     & 75.         & -2.6168    & 49.931     & 50.         & 93.00         \\ 
51 & 0.50246    & 0.42134    & 18.419     & 50.         & 22.349     & 51.24      & 55.902     & 66.435     \\ 
52 & 0.4572     & 0.40735    & 48.278     & 75.         & 22.349     & 51.24      & 55.902     & 66.435     \\ 
53 & 0.56315    & 0.37345    & 18.419     & 50.         & 47.315     & 52.548     & 70.711     & 48.00        \\ 
54 & 0.61793    & 0.32963    & 18.419     & 50.         & 72.28      & 53.857     & 90.139     & 36.69      \\ 
55 & 0.39381    & 0.52123    & 48.278     & 75.         & -28.891    & 73.589     & 79.057     & 111.43     \\ 
56 & 0.34264    & 0.56147    & 48.278     & 75.         & -53.857    & 72.28      & 90.139     & 126.69     \\ 
57 & 0.44388    & 0.48131    & 48.278     & 75.         & -3.9252    & 74.897     & 75.         & 93.00         \\ 
58 & 0.49196    & 0.4425     & 48.278     & 75.         & 21.041     & 76.206     & 79.057     & 74.565     \\ 
\bottomrule
\end{tabular}
\end{table*}

\begin{table*}[tb]
\sisetup{round-mode=places}
\centering
\figskip \caption{Mean color-concept association ratings for all UW-58 colors and fruit concepts (scale from 0 to 1). This table is reproduced from \cite{rathore2020}}
\label{table:assocData}
\begin{tabular}
{c*{1}{
    S[round-precision=2]
    S[round-precision=2]
    S[round-precision=2]
    S[round-precision=2]
    S[round-precision=2]
    S[round-precision=2]
    S[round-precision=2]
    S[round-precision=2]
    S[round-precision=2]
    S[round-precision=2]
    S[round-precision=2]
    S[round-precision=2]
}}
\toprule
\multicolumn{1}{c}{\textbf{Color}} &
\multicolumn{1}{c}{\textbf{Mango}} & \multicolumn{1}{c}{\textbf{Waterm.}} & \multicolumn{1}{c}{\textbf{Honeyd.}} & \multicolumn{1}{c}{\textbf{Cantal.}} & \multicolumn{1}{c}{\textbf{Grapefr.}} & \multicolumn{1}{c}{\textbf{Strawb.}} & \multicolumn{1}{c}{\textbf{Raspb.}} & \multicolumn{1}{c}{\textbf{Blueb.}} & \multicolumn{1}{c}{\textbf{Avocado}} &
\multicolumn{1}{c}{\textbf{Orange}} &
\multicolumn{1}{c}{\textbf{Lime}} &
\multicolumn{1}{c}{\textbf{Lemon}} \\ 
\midrule
1 & 0.111435185 & 0.092314815 & 0.121666667 & 0.092037037 & 0.091759259 & 0.100787037 & 0.126527778 & 0.512222222 & 0.111944444 & 0.09712963 & 0.108009259 & 0.12087963 \\ 
2 & 0.076203704 & 0.067962963 & 0.106203704 & 0.094490741 & 0.083564815 & 0.082268519 & 0.117685185 & 0.808055556 & 0.116805556 & 0.07837963 & 0.109907407 & 0.075185185 \\ 
3 & 0.107685185 & 0.129814815 & 0.128981481 & 0.101944444 & 0.127222222 & 0.094907407 & 0.106851852 & 0.596435185 & 0.101296296 & 0.080555556 & 0.106064815 & 0.084027778 \\ 
4 & 0.108287037 & 0.11287037 & 0.127037037 & 0.070555556 & 0.115555556 & 0.1025 & 0.093657407 & 0.397916667 & 0.104398148 & 0.102731481 & 0.098981481 & 0.127175926 \\ 
5 & 0.102314815 & 0.101527778 & 0.084768519 & 0.094537037 & 0.186111111 & 0.100833333 & 0.220833333 & 0.278240741 & 0.100185185 & 0.092083333 & 0.074027778 & 0.076296296 \\ 
6 & 0.115277778 & 0.185231481 & 0.098842593 & 0.133472222 & 0.230462963 & 0.218101852 & 0.3 & 0.2175 & 0.130925926 & 0.100462963 & 0.0775 & 0.107175926 \\ 
7 & 0.108888889 & 0.212083333 & 0.136435185 & 0.151666667 & 0.253333333 & 0.269814815 & 0.282916667 & 0.129027778 & 0.13337963 & 0.14 & 0.089444444 & 0.120324074 \\ 
8 & 0.111435185 & 0.252685185 & 0.093425926 & 0.102592593 & 0.230046296 & 0.318703704 & 0.484259259 & 0.179722222 & 0.066574074 & 0.114259259 & 0.069259259 & 0.080694444 \\ 
9 & 0.16 & 0.38337963 & 0.098611111 & 0.129074074 & 0.36537037 & 0.355138889 & 0.430231481 & 0.136018519 & 0.07162037 & 0.121851852 & 0.090833333 & 0.089537037 \\ 
10 & 0.148611111 & 0.400972222 & 0.112083333 & 0.094305556 & 0.297268519 & 0.417222222 & 0.473888889 & 0.108564815 & 0.078703704 & 0.151342593 & 0.067546296 & 0.090648148 \\ 
11 & 0.103564815 & 0.076296296 & 0.094953704 & 0.086666667 & 0.098888889 & 0.069583333 & 0.136805556 & 0.69712963 & 0.09212963 & 0.093101852 & 0.093888889 & 0.097083333 \\ 
12 & 0.095555556 & 0.075648148 & 0.093287037 & 0.078009259 & 0.089166667 & 0.062962963 & 0.152916667 & 0.861666667 & 0.113564815 & 0.09587963 & 0.071481481 & 0.090833333 \\ 
13 & 0.080277778 & 0.089537037 & 0.120277778 & 0.096851852 & 0.13625 & 0.10587963 & 0.149444444 & 0.418611111 & 0.119259259 & 0.090046296 & 0.100787037 & 0.09037037 \\  
14 & 0.068240741 & 0.088842593 & 0.100185185 & 0.081203704 & 0.161527778 & 0.133611111 & 0.20037037 & 0.283472222 & 0.109907407 & 0.083472222 & 0.076990741 & 0.09125 \\ 
15 & 0.144768519 & 0.201944444 & 0.114722222 & 0.097175926 & 0.25912037 & 0.260740741 & 0.328518519 & 0.185833333 & 0.089212963 & 0.093148148 & 0.091666667 & 0.104907407 \\ 
16 & 0.124814815 & 0.283148148 & 0.103333333 & 0.107685185 & 0.223888889 & 0.327638889 & 0.419537037 & 0.143101852 & 0.078287037 & 0.133194444 & 0.073101852 & 0.092361111 \\ 
17 & 0.092777778 & 0.088101852 & 0.109537037 & 0.081018519 & 0.107314815 & 0.074074074 & 0.126018519 & 0.737824074 & 0.088842593 & 0.087546296 & 0.110324074 & 0.095972222 \\ 
18 & 0.082361111 & 0.101342593 & 0.115694444 & 0.0725 & 0.108009259 & 0.065092593 & 0.12162037 & 0.851527778 & 0.11162037 & 0.086574074 & 0.086018519 & 0.087453704 \\ 
19 & 0.093564815 & 0.118009259 & 0.107222222 & 0.092962963 & 0.155231481 & 0.104398148 & 0.153287037 & 0.411157407 & 0.104675926 & 0.094027778 & 0.080648148 & 0.079351852 \\ 
20 & 0.093194444 & 0.144398148 & 0.116574074 & 0.092453704 & 0.2025 & 0.218518519 & 0.294814815 & 0.190046296 & 0.115648148 & 0.122824074 & 0.072037037 & 0.084027778 \\ 
21 & 0.114398148 & 0.150416667 & 0.19625 & 0.123564815 & 0.095046296 & 0.09375 & 0.100972222 & 0.320925926 & 0.207546296 & 0.089166667 & 0.206157407 & 0.079351852 \\ 
22 & 0.110092593 & 0.116990741 & 0.158518519 & 0.112824074 & 0.103703704 & 0.12037037 & 0.153611111 & 0.310925926 & 0.148564815 & 0.117685185 & 0.165648148 & 0.11125 \\ 
23 & 0.050601852 & 0.143240741 & 0.061944444 & 0.072314815 & 0.053888889 & 0.110138889 & 0.082037037 & 0.111851852 & 0.132268519 & 0.051481481 & 0.043564815 & 0.055694444 \\ 
24 & 0.066296296 & 0.101018519 & 0.101435185 & 0.078888889 & 0.078657407 & 0.08287037 & 0.068842593 & 0.126157407 & 0.102592593 & 0.083009259 & 0.072731481 & 0.068657407 \\ 
25 & 0.076805556 & 0.089166667 & 0.122777778 & 0.12462963 & 0.077222222 & 0.075601852 & 0.093796296 & 0.107083333 & 0.116759259 & 0.072407407 & 0.073101852 & 0.108842593 \\ 
26 & 0.143333333 & 0.160601852 & 0.115231481 & 0.152453704 & 0.107685185 & 0.094259259 & 0.088564815 & 0.11337963 & 0.09587963 & 0.102916667 & 0.118564815 & 0.152361111 \\ 
27 & 0.100138889 & 0.146064815 & 0.143240741 & 0.099537037 & 0.105648148 & 0.093472222 & 0.084259259 & 0.121990741 & 0.105092593 & 0.106111111 & 0.146388889 & 0.157037037 \\ 
28 & 0.112175926 & 0.326527778 & 0.083935185 & 0.139305556 & 0.269074074 & 0.450231481 & 0.512268519 & 0.168888889 & 0.074212963 & 0.122916667 & 0.067268519 & 0.080092593 \\ 
29 & 0.222592593 & 0.51625 & 0.155787037 & 0.225324074 & 0.514490741 & 0.470694444 & 0.456712963 & 0.125092593 & 0.089768519 & 0.176666667 & 0.085648148 & 0.106759259 \\ 
30 & 0.207361111 & 0.380416667 & 0.159583333 & 0.216388889 & 0.457962963 & 0.399444444 & 0.366574074 & 0.113194444 & 0.097824074 & 0.183842593 & 0.098657407 & 0.130324074 \\ 
31 & 0.185972222 & 0.635277778 & 0.145 & 0.186342593 & 0.568009259 & 0.587824074 & 0.564537037 & 0.125740741 & 0.075740741 & 0.184814815 & 0.067175926 & 0.08912037 \\ 
32 & 0.151898148 & 0.523935185 & 0.108101852 & 0.125462963 & 0.340972222 & 0.579953704 & 0.573194444 & 0.12412037 & 0.071527778 & 0.165092593 & 0.074675926 & 0.089212963 \\ 
33 & 0.177824074 & 0.72375 & 0.313425926 & 0.209398148 & 0.138055556 & 0.356203704 & 0.147407407 & 0.137268519 & 0.835555556 & 0.126990741 & 0.710648148 & 0.150925926 \\ 
34 & 0.254305556 & 0.531435185 & 0.48337963 & 0.267916667 & 0.138611111 & 0.291157407 & 0.161435185 & 0.13712963 & 0.689953704 & 0.129490741 & 0.664212963 & 0.218240741 \\ 
35 & 0.166157407 & 0.319675926 & 0.561712963 & 0.316481481 & 0.150925926 & 0.188611111 & 0.137453704 & 0.139490741 & 0.404351852 & 0.148333333 & 0.457268519 & 0.25287037 \\ 
36 & 0.153888889 & 0.275833333 & 0.362638889 & 0.205416667 & 0.133935185 & 0.156944444 & 0.129351852 & 0.181481481 & 0.32162037 & 0.119027778 & 0.39212963 & 0.151944444 \\ 
37 & 0.157037037 & 0.18287037 & 0.238009259 & 0.209212963 & 0.148333333 & 0.115787037 & 0.082222222 & 0.142314815 & 0.460740741 & 0.143796296 & 0.232777778 & 0.157222222 \\ 
38 & 0.276851852 & 0.149027778 & 0.291990741 & 0.369351852 & 0.192638889 & 0.114907407 & 0.095787037 & 0.109675926 & 0.373333333 & 0.209490741 & 0.186018519 & 0.320833333 \\ 
39 & 0.338009259 & 0.150833333 & 0.357453704 & 0.424861111 & 0.256990741 & 0.138564815 & 0.108935185 & 0.123009259 & 0.161296296 & 0.291111111 & 0.185231481 & 0.375648148 \\ 
40 & 0.274537037 & 0.126527778 & 0.17087963 & 0.285972222 & 0.181018519 & 0.151759259 & 0.13162037 & 0.088657407 & 0.156944444 & 0.444953704 & 0.076805556 & 0.152685185 \\ 
41 & 0.542824074 & 0.20712963 & 0.253425926 & 0.557037037 & 0.559537037 & 0.218101852 & 0.177592593 & 0.107268519 & 0.105648148 & 0.53 & 0.128611111 & 0.197407407 \\ 
42 & 0.428472222 & 0.269305556 & 0.293148148 & 0.569212963 & 0.582314815 & 0.245462963 & 0.228796296 & 0.102962963 & 0.097592593 & 0.417731481 & 0.136435185 & 0.197824074 \\ 
43 & 0.418703704 & 0.528009259 & 0.181342593 & 0.314768519 & 0.634953704 & 0.472824074 & 0.418564815 & 0.087546296 & 0.089027778 & 0.396574074 & 0.114166667 & 0.143796296 \\ 
44 & 0.174351852 & 0.741111111 & 0.12962963 & 0.132037037 & 0.450416667 & 0.750833333 & 0.640509259 & 0.085555556 & 0.081898148 & 0.217083333 & 0.08337963 & 0.102731481 \\ 
45 & 0.277268519 & 0.51875 & 0.428888889 & 0.292175926 & 0.124907407 & 0.292222222 & 0.13375 & 0.111296296 & 0.73125 & 0.122361111 & 0.72537037 & 0.26212963 \\ 
46 & 0.274259259 & 0.361111111 & 0.612962963 & 0.279027778 & 0.177407407 & 0.194861111 & 0.13912037 & 0.118935185 & 0.485092593 & 0.16337963 & 0.572175926 & 0.417962963 \\ 
47 & 0.226666667 & 0.632175926 & 0.470972222 & 0.252314815 & 0.131759259 & 0.325509259 & 0.149027778 & 0.129953704 & 0.690694444 & 0.100185185 & 0.825833333 & 0.223333333 \\ 
48 & 0.221435185 & 0.539861111 & 0.523009259 & 0.268935185 & 0.134166667 & 0.255277778 & 0.165 & 0.157731481 & 0.531064815 & 0.131157407 & 0.710138889 & 0.218842593 \\ 
49 & 0.38037037 & 0.137824074 & 0.297962963 & 0.389722222 & 0.248518519 & 0.097824074 & 0.104212963 & 0.10712963 & 0.317453704 & 0.281759259 & 0.210185185 & 0.535462963 \\ 
50 & 0.535231481 & 0.116527778 & 0.431666667 & 0.540648148 & 0.365694444 & 0.11962963 & 0.112546296 & 0.111805556 & 0.208564815 & 0.383564815 & 0.16912037 & 0.630138889 \\ 
51 & 0.751805556 & 0.092407407 & 0.297083333 & 0.692824074 & 0.489861111 & 0.118009259 & 0.09787037 & 0.113055556 & 0.123287037 & 0.849722222 & 0.13537037 & 0.269166667 \\ 
52 & 0.618101852 & 0.14337963 & 0.379398148 & 0.780324074 & 0.578611111 & 0.163657407 & 0.190787037 & 0.086944444 & 0.119675926 & 0.648009259 & 0.147731481 & 0.332685185 \\ 
53 & 0.670231481 & 0.128935185 & 0.220462963 & 0.486574074 & 0.414675926 & 0.175694444 & 0.133333333 & 0.075 & 0.081851852 & 0.859490741 & 0.107175926 & 0.217268519 \\ 
54 & 0.315833333 & 0.680972222 & 0.113888889 & 0.215694444 & 0.325277778 & 0.745694444 & 0.559907407 & 0.065324074 & 0.065416667 & 0.348981481 & 0.090092593 & 0.143657407 \\ 
55 & 0.262083333 & 0.268935185 & 0.375277778 & 0.24412037 & 0.204398148 & 0.151527778 & 0.111018519 & 0.097916667 & 0.365555556 & 0.170555556 & 0.506342593 & 0.601435185 \\ 
56 & 0.197175926 & 0.511296296 & 0.446574074 & 0.266759259 & 0.125277778 & 0.25625 & 0.130694444 & 0.114398148 & 0.529814815 & 0.12125 & 0.773842593 & 0.253101852 \\ 
57 & 0.599814815 & 0.112268519 & 0.324814815 & 0.449444444 & 0.323194444 & 0.108287037 & 0.093518519 & 0.096759259 & 0.15037037 & 0.424444444 & 0.251527778 & 0.894444444 \\ 
58 & 0.854212963 & 0.115138889 & 0.31587963 & 0.725740741 & 0.454166667 & 0.108564815 & 0.110138889 & 0.072638889 & 0.11162037 & 0.823194444 & 0.15962963 & 0.464953704 \\ 
\bottomrule
\end{tabular}
\end{table*}

\end{document}